\documentclass[iop,numberedappendix]{emulateapj}
\usepackage{natbib}
\usepackage{multirow}
\usepackage{upgreek}
\usepackage[colorlinks=true, pdfstartview=FitV, linkcolor=blue,citecolor=blue, urlcolor=blue]{hyperref}
\usepackage{wasysym}
\usepackage{txfonts}
\usepackage{gensymb}
\usepackage{morefloats}
\usepackage{threeparttable}
\usepackage{rotating}
\usepackage{graphicx}
\slugcomment{version \today}

\slugcomment{version \today: fm}
\shorttitle{Unidentified Gamma-ray Sources IV}
\shortauthors{A. Paggi et al.}

\begin{document}
\title{Unveiling the nature of the unidentified gamma-ray sources IV:\\the \textit{Swift} catalog of potential X-ray counterparts}
\author{
A. Paggi\altaffilmark{1}, 
F. Massaro\altaffilmark{2}, 
R. D'Abrusco\altaffilmark{1}, 
H. A. Smith\altaffilmark{1},
N. Masetti\altaffilmark{3},
M. Giroletti\altaffilmark{4},
G. Tosti\altaffilmark{5},
S. Funk \altaffilmark{2}
}

\altaffiltext{1}{Harvard - Smithsonian Astrophysical Observatory, 60 Garden Street, Cambridge, MA 02138, USA}
\altaffiltext{2}{SLAC National Laboratory and Kavli Institute for Particle Astrophysics and Cosmology, 2575 Sand Hill Road, Menlo Park, CA 94025, USA}
\altaffiltext{3}{INAF - Istituto di Astrofisica Spaziale e Fisica Cosmica di Bologna, via Gobetti 101, 40129, Bologna, Italy}
\altaffiltext{4}{INAF Istituto di Radioastronomia, via Gobetti 101, 40129, Bologna, Italy}
\altaffiltext{5}{Dipartimento di Fisica, Universit\`a degli Studi di Perugia, 06123 Perugia, Italy}

\begin{abstract}
A significant fraction (\(\sim 30\)\%) of the high-energy gamma-ray sources 
listed in the second \textit{Fermi} LAT (2FGL) catalog are still of unknown 
origin, being not yet associated with counterparts at lower energies. In 
order to investigate the nature of these enigmatic sources, we present here 
an extensive search of X-ray {sources lying in the positional 
uncertainty region} of a selected sample of these Unidentified Gamma-ray 
Sources (UGSs) that makes use of all available observations performed by 
the \textit{Swift} X-ray Telescope before March 31, 2013, available for 205 
UGSs. To detect the fainter sources, we merged all the observations 
covering the {\textit{Fermi} LAT} positional uncertainty region at 
95\% level of confidence of each UGSs. This yields a catalog of 357 X-ray 
sources, finding {candidate} X-ray counterparts for \(\sim 70\)\% of 
the selected sample. In particular, 25\% of the UGSs feature a single X-ray 
source within their positional uncertainty region while 45\% have multiple 
X-ray sources. For each X-ray source we also looked in the corresponding 
\textit{Swift} UVOT merged images for optical and ultraviolet counterparts, 
also performing source photometry. We found ultraviolet-optical 
correspondences for \(\sim 70\)\% of the X-ray sources. We searched several 
major radio, infrared, optical and ultraviolet surveys for possible 
counterparts within the positional error of the sources in the X-ray 
catalog to obtain additional information on their nature. Applying the 
kernel density estimator technique to infrared colors of WISE counterparts 
of our X-ray sources we select 6 \(\gamma\)-ray blazar candidates. In 
addition, comparing our results with previous analyses, we select 11 
additional \(\gamma\)-ray blazar candidates.
\end{abstract}


\section{Introduction}
\label{sec:intro}

One of the biggest challenges of modern \(\gamma\)-ray astronomy and one of 
the main scientific objectives of the ongoing \textit{Fermi} mission is 
unraveling the nature of the Unidentified Gamma-ray Sources (UGSs) 
\citep[e.g.,][]{abdo09,2009ApJ...697.1071A}. 

Since the Third EGRET catalog 
(3EG)\footnote{\href{http://heasarc.gsfc.nasa.gov/W3Browse/cgro/egret3.html}
{http://heasarc.gsfc.nasa.gov/W3Browse/cgro/egret3.html}} 
\citep[e.g.,][]{hartman99} the fraction of \(\gamma\)-ray sources without 
an assigned counterpart at low energies has been significant \(\sim 30\)\% 
\citep[e.g.,][]{2003ApJ...590..109S}.  This situation was mostly unchanged 
in the revised EGRET catalog \citep[EGR;][]{2008A&A...489..849C}, even 
though the improved background modeling applied in the EGR resulted in 
fewer \(\gamma\)-ray detections (188 sources in total, in contrast to 271 
listed in 3EG); 87 out of 188 EGR entries remain unassociated.

The UGSs at low Galactic latitude (\(|b|<10 \degr\)) are expected to be 
associated with local objects lying in our Galaxy, such as molecular clouds 
(as consequence of interaction with cosmic-rays), supernova remnants, 
massive stars, pulsars and pulsar wind nebulae, or X-ray binaries \citep[see,e.g.,][]{1999APh....11..277G,2010PASJ...62..769C,2012ApJ...745..140Y,2013Sci...339..807A,2013A&A...553A..34D} although there are few rare 
cases of \(\gamma\)-ray blazars detected through the Galactic plane 
\citep[e.g. Fermi J0109+6134, see][]{2010ApJ...718L.166V}. On the other 
hand, the population of UGSs above the Galactic plane is generally believed 
to be dominated by extragalactic sources, although there is a suspected 
Galactic component as well 
\citep[e.g.,][]{1996ApJ...463..105O,2000ApJ...541..180M,2001ASSL..267...17R,nolan12}. According to one of the most recent \textit{Fermi} discoveries, 
several millisecond pulsars have been found at high Galactic latitudes 
\citep{2010ApJ...712.1209A,2010ApJ...712..957A,nolan12}.

A large fraction of these UGSs could be blazars, the rarest class of 
radio-loud active galactic nuclei, whose emission dominates the gamma-ray 
sky \citep[e.g.,][]{mukherjee97,abdo10}. Their observational properties are 
generally interpreted in terms of a relativistic jet aligned within a small 
angle to our line of sight \citep{1978bllo.conf..328B}.

The blazar spectral energy distributions (SEDs) typically show two peaks. 
The first one, lying in the range of {radio} - soft X-rays, is widely held 
to be due to synchrotron emission by highly relativistic electrons within 
their jet. The second one lies at hard X-ray or \(\gamma\)-ray energies, 
and is interpreted as inverse Compton upscattering by the same electrons of 
the seed photons provided by the synchrotron emission 
\citep{1996ApJ...463..555I,2008ApJ...686..181F} with the possible addition 
of seed photons from outside the jets yielding contributions to the 
non-thermal radiations due to external inverse Compton scattering 
\citep[see][]{1993ApJ...416..458D,2002ApJ...575..667D,2009ApJ...692...32D,2013ApJ...763..134F} often dominating their \(\gamma\)-ray outputs 
\citep{ackermann11}.

Blazars are also know X-ray sources since \textit{ROSAT} DXRBS 
\citep{1998AJ....115.1253P,2001MNRAS.323..757L} and
\textit{Einstein} IPC \citep{1992ApJS...80..257E,1999AAS...195.1601P} 
surveys \citep[see also][]{2000AIPC..515...53P}. Since then, the X-ray 
properties of blazars have been deeply investigated by many authors 
\citep[see for example][]{1994MNRAS.268L..51G,1995ApJ...444..567P,2011ApJ...739...73M,2011ApJ...742L..32M}. \citet{2008A&A...489.1047M} in particular studied 
\textit{Swift} observations of a sample of low and intermediate peaked BL 
Lacs, for which the X-ray emission is expected to lie in the ``valley" 
between the low and high energy spectral components, finding these sources 
to be bright in the X-ray with fluxes above \(\sim {10}^{-13}\mbox{ 
erg}\mbox{ cm}^{-2}\mbox{ s}^{-1}\). In addition we note that \(\sim 75\%\) 
of the \(\gamma\)-ray blazars listed in the Second LAT AGN Catalog 
\citep[2LAC,][]{ackermann11} are also X-ray sources with fluxes above 
\(\sim {10}^{-14}\mbox{ erg}\mbox{ cm}^{-2}\mbox{ s}^{-1}\).

However, due to the incompleteness of the current radio and X-ray surveys 
used for the gamma-ray associations, it is not always possible to 
identify a blazar-like counterpart to a UGS\footnote{{We note that, 
in the following, we will refer to a source lying into the positional 
uncertainty region of a \(\gamma\)-ray source as ``candidate counterpart", 
while we will use the term ``blazar candidate" for the \(\gamma\)-ray 
source together with its unique blazar-like counterpart.}}.

Radio follow up observations of UGSs have been performed or are still in 
progress 
\citep[e.g.,][]{kovalev09a,kovalev09b,2010ApJ...718..587M,2013MNRAS.432.1294P}. \citet{massaro2013b} recently proposed a method for searching 
\(\gamma\)-ray blazar-like {candidate} counterparts of the UGSs 
based on the combination of radio observations from Westerbork Northern Sky 
Survey \citep[WENSS;][]{1997A&AS..124..259R}, those of the NRAO Very Large 
Array Sky survey \citep[NVSS;][]{1998AJ....115.1693C} and the Very Large 
Array Faint Images of the Radio Sky at Twenty-Centimeters 
\citep[FIRST,][]{1995ApJ...450..559B,white97}.
  
In addition, a procedure to recognize blazar-like {candidate} 
counterparts for UGSs on the basis of their infrared (IR) colors have been 
successfully implemented by \citet{paper2,2013ApJS..206...12D} and 
\citet{paper3,2013arXiv1303.3585M} making use of the Wide-Field Infrared 
Survey Explorer (WISE) all-sky data \citep{cutri12a}. WISE data also proven 
to be useful to address the widely entertained field of mid-infrared AGN 
selection (\citealt{2005ApJ...631..163S,2012ApJ...753...30S}, see also 
\citealt{2010ApJ...708..584E,2010ApJ...717.1181P}).

Additional attempts have been recently developed to associate or to 
characterize the UGSs using pointed \textit{Swift} observations 
\citep[e.g.,][]{mirabal09a,mirabal09b,2012ApJ...757..176K}, and/or with 
several statistical approaches \citep[e.g.,][]{mirabal10,ackermann12}. 
Moreover, in the last two years the \textit{Chandra} and \textit{Suzaku} 
X-ray telescopes have been used to investigate the nature of the UGSs 
\citep[e.g.,][]{2011PASJ...63S.857F,2011ApJ...729..103M,2011PASJ...63S.873M,2012ApJ...756...33C,2012PASJ...64..112M}.

The characterization of X-ray emission from UGSs is of particular interest. 
All \(\gamma\)-ray sources associated in the second \textit{Fermi} LAT 
(2FGL) catalog have a clear radio counterpart \citep{nolan12} leading to 
the so called radio-\(\gamma\)-ray connection in the case of blazars 
\citep[e.g.,][]{ghirlanda10,ackermann11,massaro2013b}. However this is not 
the case for the X-ray sources. It is not clear at the moment if all 
\(\gamma\)-ray sources feature an X-ray counterpart and therefore a 
systematic study of X-ray emission from UGS is useful to investigate their 
nature.

Motivated by these researches, we investigate the X-ray-\(\gamma\) 
connection presenting in this paper a catalog of X-ray {sources 
lying in the positional uncertainty region of} all UGSs listed in 2FGL 
without any \(\gamma\)-ray analysis flag, making use of all available 
observations performed by \textit{Swift} X-ray Telescope (XRT) up to March 
31, 2013, and we investigate their multi-wavelength properties.

For {X-ray} sources with a WISE counterpart we then apply the Kernel 
Density Estimation (KDE) technique to compare their IR colors to those of 
known \(\gamma\)-ray blazars, selecting 44 new blazar-like 
{candidate} counterparts and 6 \(\gamma\)-ray blazars candidates as 
a result.

The paper is organized as follows: Section \ref{sec:sample} is devoted to 
the UGS sample definition while Section \ref{sec:swift} describes the main 
data reduction procedure adopted for the \textit{Swift} XRT and 
\textit{Swift} UVOT observations. The complete list of X-ray sources that 
could be potential counterpart of UGSs in the 2FGL catalog is presented in 
Section \ref{sec:ugs}. In Section \ref{sec:kde} we illustrate our selection 
of new \(\gamma\)-ray blazar candidates. In Section \ref{sec:comparison} we 
compare our results with different, previous selections, and Section 
\ref{sec:summary} is dedicated to our conclusions.

\section{Sample selection}\label{sec:sample}

The initial sample considered in our analysis is constituted by the 299 
UGSs in the 2FGL catalog that do not present any \(\gamma\)-ray analysis 
flag\footnote{Analysis flags in 2FGL identify a number of conditions that 
can shed doubt on a source, and they are described in detail in Table 3 of 
\citet{nolan12}.} \citep{nolan12}.

Up to March 31, 2013, 205 of these sources feature at least one X-ray 
observation in the \textit{Swift} master catalog\footnote{\href{http://heasarc.gsfc.nasa.gov/W3Browse/all/swiftmastr.html}{http://heasarc.gsfc.nasa.gov/W3Browse/all/swiftmastr.html}} 
performed in photon counting (PC) mode, and covering the positional 
uncertainty region at 95\% level of confidence as reported in the 2FGL. The final sample considered in this analysis is therefore 
constituted by the above selected 205 sources.

\begin{figure}
\begin{center}
\includegraphics[scale=0.3]{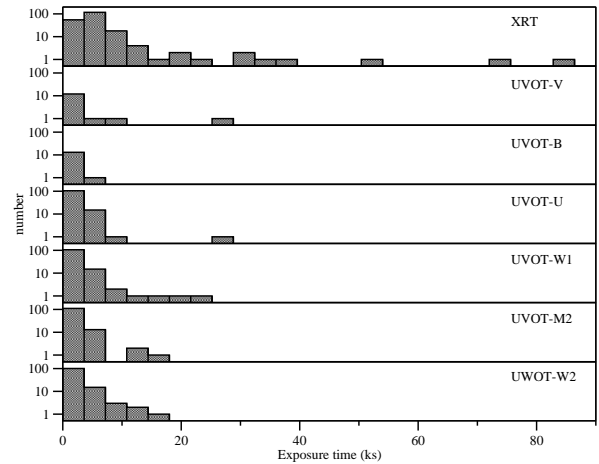}
\end{center}
\caption{Histograms of total exposures of the merged observations discussed in Section \ref{sec:sample}.}\label{exposures}
\end{figure}

The \textit{Swift} observations have variable exposures, and to detect the 
fainter X-ray objects we merged all the observations corresponding to each 
UGSs (see Section \ref{sec:swift} for details on the reduction procedures), 
obtaining the total exposures shown in Figure \ref{exposures}.

\section{\textit{Swift} observations and data reduction procedures}
\label{sec:swift}

\textit{Swift} has proven to be an excellent multi-frequency observatory for blazar research, so far observing 
hundreds of sources \citep[e.g.,][]{2007SPIE.6688E..14M,2012A&A...548A..87M,2012AAS...21941506D} and yielding 
an extremely rich and unique database of multi-frequency (optical, UV, X-ray), simultaneous blazar observations. 
Several papers on samples selected with different criteria have already been published, including: blazars detected at 
TeV energies \citep[e.g.,][]{2008A&A...478..395M,2011A&A...528C...1M,2011ApJ...739...73M,2011ApJ...742L..32M}, 
simultaneous optical-to-X-ray observations of flaring TeV sources 
\citep[e.g.,][]{2007A&A...462..889P,2007A&A...467..501T} as well as the investigation of low and high frequency 
peaked BL Lacs \citep[e.g.,][]{2010A&A...512A..74M,2012A&A...541A.160G}. \textit{Swift} has also been used for UV-optical 
and X-ray follow-up observations of TeV flaring blazars \citep[e.g.,][]{2011ApJ...742..127A,2012A&A...544A.142A,2013A&A...552A.118H}
and has also been useful in obtaining photometric redshift constraints for many \textit{Fermi}-detected BL Lacs \citep{2012A&A...538A..26R}.

Once \textit{Fermi} was launched, the \textit{Swift} XRT Survey of 
\textit{Fermi} Unassociated Sources was started to perform follow-up 
observations of the UGSs in an attempt to find their potential X-ray 
counterparts\footnote{\href{http://www.swift.psu.edu/unassociated/}{http://www.swift.psu.edu/unassociated/}} (PI A. Falcone). In the following sections 
we analyze all the data collected between the beginning of the follow-up 
program until March 31, 2013, for the selected sample of UGSs described in 
Section \ref{sec:sample}.

During these observations, \textit{Swift} operated with all its instruments in data taking mode. For our analysis, 
however, we consider only \textit{Swift} XRT \citep{2005SSRv..120..165B} and \textit{Swift} UVOT \citep{2005SSRv..120...95R} data.

\subsection{\textit{Swift} XRT data reduction}
\label{sec:reduction}

The XRT data were processed using the XRTDAS software \citep{capalbi2005} developed at the ASI Science Data 
Center and included in the HEAsoft package (v. 6.13) distributed by HEASARC.
For each observation of the sample, calibrated and cleaned PC mode event files were produced with the 
\textsc{xrtpipeline} task (ver. 0.12.6), producing exposure maps for each observation. In addition to the screening 
criteria used by the standard pipeline processing, we applied a further filter to screen background spikes that can 
occur when the angle between the pointing direction of the satellite and the bright Earth limb is low. In order to 
eliminate this so called bright Earth effect, due to the scattered optical light that usually occurs towards the beginning 
or the end of each orbit, we used the procedure proposed by \citet{2011A&A...528A.122P} and 
\citet{2013A&A...551A.142D}. We monitored the count rate on the CCD border and, through the \textsc{xselect} 
package, we excluded time intervals when the count rate in this region exceeded 40 counts/s; moreover, we selected 
only time intervals with CCD temperatures less than \(-50\degree\mbox{C}\) (instead of the standard limit of 
\(-47\degree\mbox{C}\)) since contamination by dark current and hot pixels, which increase the low energy background, is strongly temperature dependent \citep{2013A&A...551A.142D}.

We then proceeded to merge cleaned event files obtained with this procedure using \textsc{xselect}, considering only 
observations with telescope aim point falling in a circular region of 12' radius centered in the median of the 
individual aim points, in order to have a uniform exposure. The corresponding merged exposure maps were 
then generated by summing the exposure maps of the individual observations with \textsc{ximage} (ver. 4.5.1).

\subsection{\textit{Swift} XRT source detection}
\label{sec:xrt}

To detect X-ray sources in the merged XRT images, we made used of the \textsc{ximage} detection algorithm \textsc{detect}, 
which locates the point sources using a sliding-cell method. The average background intensity is estimated in several 
small square boxes uniformly located within the image. The position and intensity of each detected source are 
calculated in a box whose size maximizes the signal-to-noise ratio. The net counts are corrected for dead times and 
vignetting using the appropriate exposure maps, and for the fraction of source counts that fall outside the box where 
the net counts are estimated, using the PSF calibration. Count rate statistical and systematic uncertainties are added 
quadratically. The algorithm was set to work in bright mode, which is recommended for crowded fields and fields 
containing bright sources, since it can reconstruct the centroids of very nearby sources.

We also evaluated the net count rates for the detected sources with the \textsc{sosta} algorithm that, besides the net 
count rates and the respective uncertainties, yields the statistical significance of each source. We note that the 
uncertainties in the count rates returned by \textsc{sosta} are purely statistical - i.e. do not include systematic errors 
- and are in general smaller than those given by \textsc{detect}. \textsc{sosta} also yields slightly different count rates 
from \textsc{detect}, which are in most cases more accurate, because \textsc{detect} uses a global background for the 
entire image, whereas \textsc{sosta} uses a local background. Thus we report both values in our analysis.

The catalog was then cleaned from spurious sources - usually occurring at count rates higher than \(0.2\mbox{ ph}
\mbox{ s}^{-1}\) - by visual inspection of all the observations. Finally, we refined the source position and relative 
positional errors by the task \textsc{xrtcentroid} of the XRTDAS package, and considered only sources falling
in a circular region of radius equal to the semi-major axis of the ellipse corresponding to the positional uncertainty 
region of the \textit{Fermi} source at 95\% level of confidence and centered at the 2FGL position of the 
\(\gamma\)-ray source (consistently with \citealt{2013arXiv1303.3585M}). The source designation we adopt for a 
source with RA HH:MM:SS.s and DEC \(\pm\)DD:MM:SS is SWXRTJHHMMSS.s\(\pm\)DDMMSS, as per 
\citet{2013A&A...551A.142D}. The results of the detection process are presented in Appendix \ref{app:tables} in Table 
\ref{table_xrt}.

\subsection{\textit{Swift} UVOT observations}
\label{sec:uvot}

We note that 203 out of the 205 UGSs that constitute our sample have been 
also observed in the optical and UV by UVOT. We then produced for each 
X-ray observation the corresponding merged UVOT event files adopting 
standard 
procedures\footnote{\href{http://www.swift.ac.uk/analysis/uvot/image.php}{http://www.swift.ac.uk/analysis/uvot/image.php}}. After checking the correct 
WCS alignment of our images with USNO-B Catalog 
\citep{2003AJ....125..984M}, we merged them with \textsc{fappend} (part of 
FTOOLS package ver. 6.13) and then merged the images with 
\textsc{uvotimsum}; the same procedure was applied to produce merged 
exposure maps.

For each X-ray source found with the procedure described in \ref{sec:xrt}, 
we looked in the corresponding UVOT images for UV-optical counterparts 
falling in the relative XRT positional error. We performed source 
photometry using the \textsc{uvotsource} task using the appropriate 
exposure map. We adopted an aperture radius of 5'', independently of the 
image filter, and took the background region in the form of circle with 
typical radius of 20'' in a source-free region of the sky 
\citep[e.g.,][]{maselli2013}.

As a comparison we also evaluated source photometry with the \textsc{uvotdetect} task, which  detects sources in UVOT 
images and extracts their count rates evaluating the background level. In general, we note that although the
\textsc{uvotsource} task yields more accurate results for extended sources, we expect to deal mostly with point-like 
sources. The results of the detection process are presented in Table \ref{table_uvot}.

\subsection{Chance coincidence probability}
\label{sec:canzo}

\begin{figure}[!t]
\begin{center}
\includegraphics[scale=0.4]{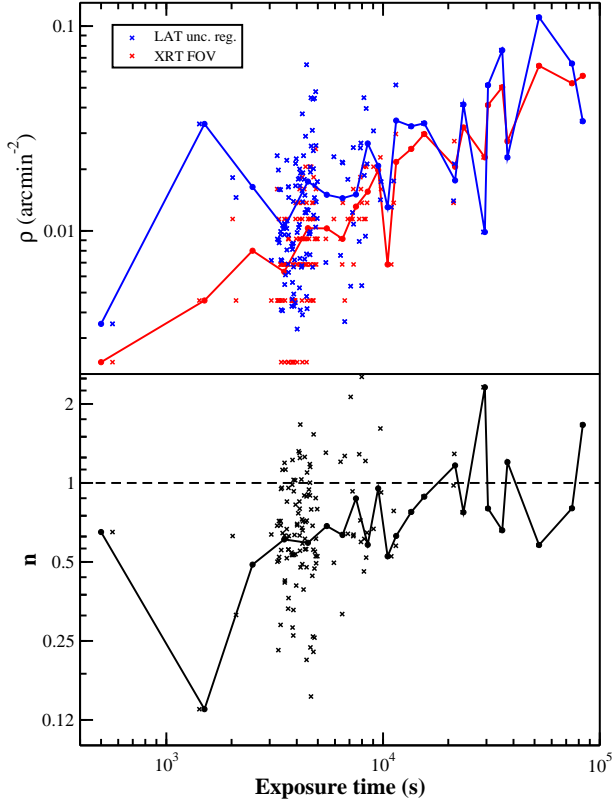}
\end{center}
\caption{(Upper panel) Mean spatial density \(\rho\) of 
X-ray sources detected inside the LAT positional uncertainty region (blue 
crosses) and in the whole \textit{Swift} XRT field of view (red crosses), 
as a function of the exposure time. With circles of the appropriate color 
we represent the average values of \(\rho\) in bins of exposure time of 1 
ks. (Lower panel) Ratio \(n\) of mean spatial density of X-ray sources 
detected in the whole \textit{Swift} XRT field of view to mean spatial 
density of X-ray sources detected inside the LAT positional uncertainty 
region, as a function of the exposure time (black crosses).With black 
circles we represent the average values of \(n\) in bins of exposure time 
of 1 ks.}\label{density}
\end{figure}

Due to considerable size of the \textit{Fermi} LAT positional 
uncertainty region {(ranging from \(\sim 2'\) to \(\sim 20'\) with 
an average size \(\sim 8'\))} several UGSs feature more than one X-ray 
source in their uncertainty region. For this reason, we performed for each 
UGS listed in Table \ref{table_xrt} simulations to evaluate the probability 
of chance coincidence detections of X-ray sources.

As a first step we evaluated the mean spatial density \(\rho\) of 
X-ray sources detected in the whole \textit{Swift} XRT field of view and 
inside the LAT positional uncertainty region. In the upper panel of Figure 
\ref{density} we present with red and blue crosses respectively these two 
densities as a function of the exposure time, while in the lower panel of 
the same figure we show with black crosses the ratio \(n\) of these two 
densities. Despite the spread, the average values of these quantities 
evaluated in bins of 1 ks (indicated with circles of the appropriate color) 
show that for exposure times higher than \(\sim 20\) ks the two mean 
densities {become} comparable.

{The mean spatial densities, however, cannot be used to properly evaluate the chance coincidence probability, since they do not take into account the spatial distribution of the X-ray sources, that is not uniform. In order to properly evaluate the chance coincidence probability we adopted a method similar to that presented by \citet{2013ApJS..206...12D}, that consists in randomly shifting the searching region (in our case, the LAT positional uncertainty region) and evaluate how many X-ray sources fall into this shifted region. For each USG listed in Table \ref{table_xrt} we generated 50 random regions of the same size of the relative LAT positional uncertainty region (and disconnected from the latter) in order to cover the whole \textit{Swift} XRT field of view. We then counted how many of these random regions contain a number of X-ray sources equal or higher than the number of X-ray sources contained inside the LAT positional uncertainty region, evaluating for each UGS the relative chance coincidence probability that, as shown in Figure \ref{density}, depends on the source exposure. We then evaluated the average chance coincidence probability for all our UGS, that is \(\sim 5\%\) with a standard deviation of \(\sim 13\%\); we can therefore conservatively evaluate a chance coincidence probability \(\lesssim 18\%\). This value makes us confident in associating the detected X-ray sources with the UGSs.}

\begin{figure}[!b]
\begin{center}
\includegraphics[scale=0.3]{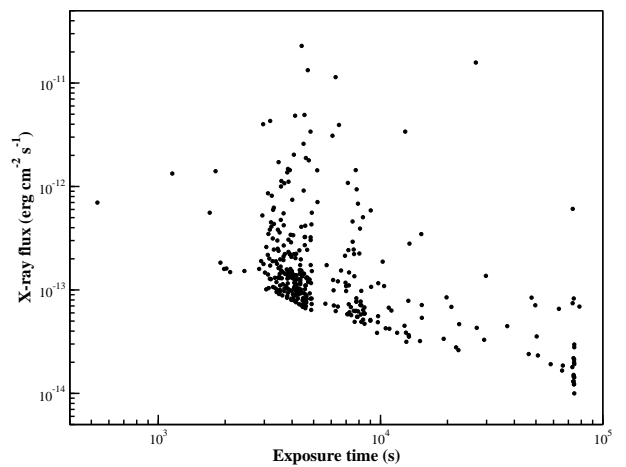}
\end{center}
\caption{Total exposure for each source of our catalog compared with the respective observed X-ray flux evaluated 
with \textsc{PIMMS} software for a powerlaw spectra with spectral index 2 and an absorption column 
density of \(5\times{10}^{20}\mbox{ cm}^{-2}\). We note that this model assumption induce an error of \(\sim 8\%\) on the estimated flux.}\label{exposures_vs_fluxes}
\end{figure}

\section{The X-ray catalog of candidate counterparts for the Unidentified Gamma-ray Sources}\label{sec:ugs}

Using the procedure described in \ref{sec:xrt}, we obtained a catalog of 
357 X-ray sources detected with a significance \(\geq 2\sigma\). In 
particular, we have 195 sources detected with a significance \(\geq 
3\sigma\), 111 sources with a significance \(\geq 4\sigma\) and 80 sources 
with a significance \(\geq 5\sigma\). We found X-ray sources consistent 
with the locations of 143 UGSs, with 51 UGSs having a single X-ray source 
and 92 UGSs having multiple X-ray sources in their positional uncertainty 
region. The remaining 62 UGSs, although overlapping with XRT-PC 
observations, do not show any X-ray counterpart.

In Figure \ref{exposures_vs_fluxes} we show for each X-ray source of our catalog the estimated X-ray flux evaluated with \textsc{PIMMS}\footnote{\href{http://heasarc.nasa.gov/docs/journal/pimms3.html}{http://heasarc.nasa.gov/docs/journal/pimms3.html}} 4.6b software for a standard powerlaw spectra with spectral index 2 and an
absorption column density fixed to \(5\times{10}^{20}\mbox{ cm}^{-2}\). Figure \ref{exposures_vs_fluxes} clearly shows the 
flux limit for an X-ray source to be detected with a specific exposure.

We searched several major radio, IR, optical and UV catalogs for possible 
counterparts within the positional errors obtained with 
\textsc{xrtcentroid} to obtain additional information on the source nature.

For the radio catalogs we considered NVSS \citep[N;][]{1998AJ....115.1693C}, Sydney University Molonglo Sky Survey 
\citep[SUMSS - S;][]{1999AJ....117.1578B,2003MNRAS.342.1117M}, FIRST \citep[F;][]{1995ApJ...450..559B} and WENSS 
\citep[W;][]{1997A&AS..124..259R} surveys. For the IR catalogs, we used the WISE \citep[w;][]{2010AJ....140.1868W} 
archival observations together with the Two Micron All Sky Survey \citep[2MASS - M;][]{2006AJ....131.1163S}
since each \textit{WISE}\ source is already associated with the closest 2MASS object by the default catalog 
\citep[see][for more details]{cutri12b}, and the UKIRT Infrared Deep Sky Survey 
\citep[UKIDSS - UK;][]{2007MNRAS.379.1599L} archival observations. For the UV catalog, we used the Galaxy Evolution 
Explorer \citep[GALEX GR6 - g;][]{2005ApJ...619L...1M} archival observations. In addition we searched for optical 
counterparts, with possible spectra available, in the Sloan Digital Sky Survey 
\citep[SDSS dr9 - s; e.g.][]{2012A&A...548A..66P} and in the Six-degree-Field Galaxy Redshift Survey 
\citep[6dFGS - 6;][]{2004MNRAS.355..747J,2009MNRAS.399..683J}. 
Finally, we searched for X-ray correspondences in the \textit{Chandra} Source Catalog \citep[CSC - C; e.g.][]{2010ApJS..189...37E}.

As anticipated in Section \ref{sec:uvot}, we cross-
checked XRT-PC observations with UVOT observations both in UV (u) and optical (o) filters. Then, we also considered 
the NASA Extragalactic Database (NED)\footnote{\href{http://ned.ipac.caltech.edu/}{http://ned.ipac.caltech.edu/}} for 
other multifrequency information. Finally, we cross correlate our sample with the USNO-B Catalog 
\citep[U;][]{2003AJ....125..984M} to identify the optical counterparts of our \(\gamma\)-ray blazar candidates; this is 
important to prepare and plan future follow up observations (see Table \ref{usnomag}).

In our catalog of 357 X-ray sources we find the following counterparts: 26 in the NVSS catalog, 6 in the SUMSS 
catalog, 5 in the FIRST catalog, 2 in the WENSS catalog, 41 in the SDSS catalog (2 with spectral observations), 5 in the 
6DFGS catalog, 194 in the USNO-B catalog, 44 in the GALEX catalog, 6 in the UKIDSS catalog, 197 in the WISE 
catalog (94 with 2MASS counterpart) and 1 in the CSC catalog. The results of this association procedure are presented in Table \ref{table_xrt} 
(column 10).

Although a proper counterpart identification would require more 
sophisticated techniques \citep[see for example][]{2006ApJ...641..140B}, 
for the scope of this work we are simply presenting a list of counterparts 
associations only based on positional match. We note that for the 
{197} X-ray sources for which we find WISE counterparts we only have 
one multiple match, while for the other catalogs considered here we have 7 
multiple matches for SDSS, 1 multiple match for GALEX, and 1 multiple match 
for UKIDSS. When multiple counterparts were found within the positional 
error we simply choose the closer one.

We add that we also checked \textit{Planck} PCCS \citep{2013arXiv1303.5088P}, Catalina CRTS \citep{2009ApJ...696..870D}, ROSAT RASS \citep{1999A&A...349..389V}, XMM-\textit{Newton} XMMMASTER \citep{2002astro.ph..6412A} and \textit{Suzaku} SUZAMASTER\footnote{\href{http://heasarc.gsfc.nasa.gov/W3Browse/all/suzamaster.html}{http://heasarc.gsfc.nasa.gov/W3Browse/all/suzamaster.html}} catalogs, finding no correspondences.

\begin{figure}
\begin{center}
\includegraphics[scale=0.5]{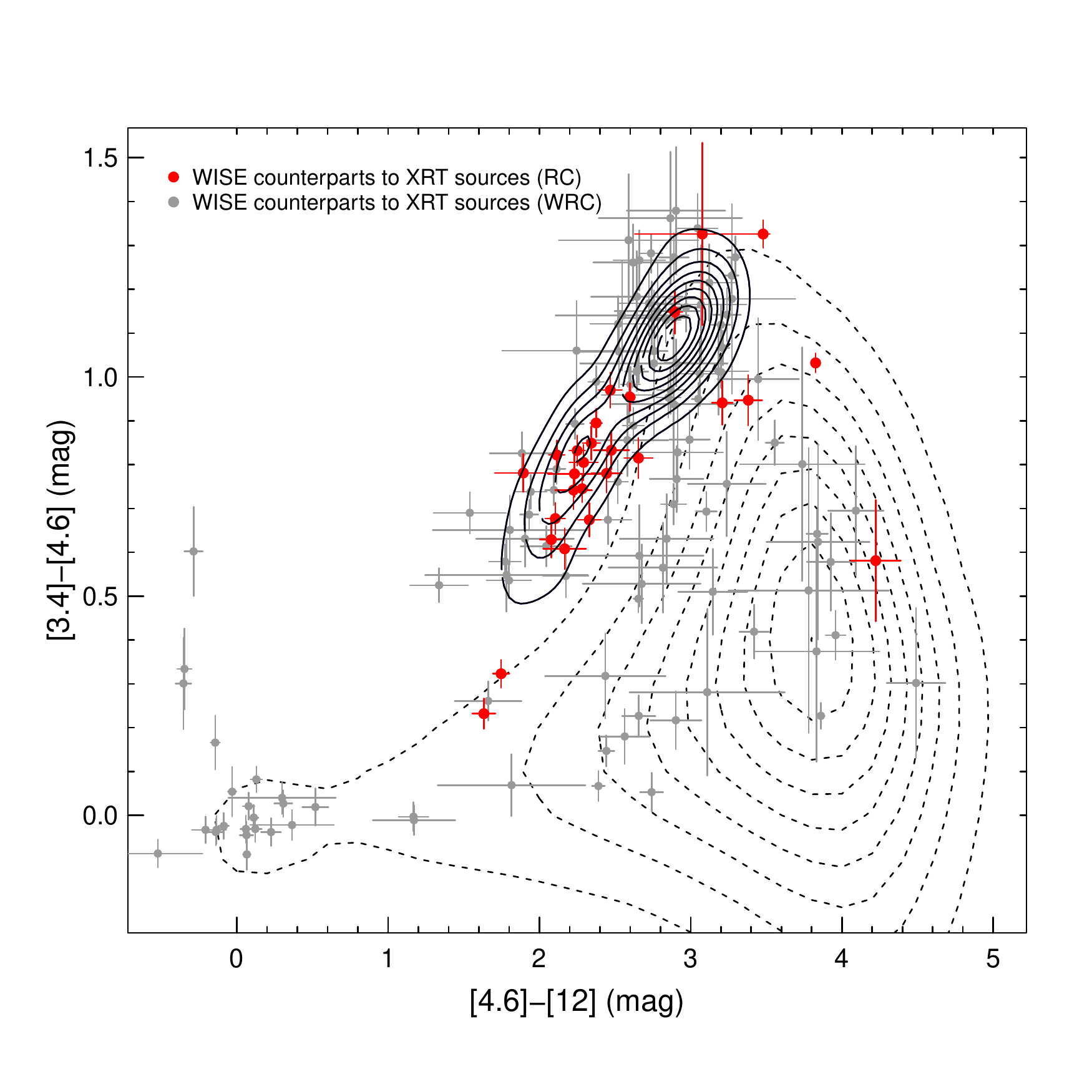}
\end{center}
\caption{Projection of the three-dimensional WISE color space on the 
two-dimensional [3.4]-[4.6] [4.6]-[12] color-color plane for XRT-PC sources 
with a WISE counterpart. {Black} lines represent the two-dimensional 
densities of WISE counterparts to know \(\gamma\)-ray blazars evaluated 
using the KDE technique, with the outermost line indicating the 90\% 
density contour normalized to the peak density. Grey circles represent 
XRT-PC sources without a radio counterpart (WRC), and red circles represent 
the XRT-PC sources with a radio counterpart (RC). Black dashed lines 
represent isodensity contours of generic WISE sources 
\citep{paper2,paper3}. The outer dashed line represent densities \(\sim 
{10}^{-4}\) times the peak density.}\label{strip}
\end{figure}

\section{Candidate \(\gamma\)-ray-blazar selection}
\label{sec:kde}

\begin{figure*}
\begin{center}
\includegraphics[scale=0.9]{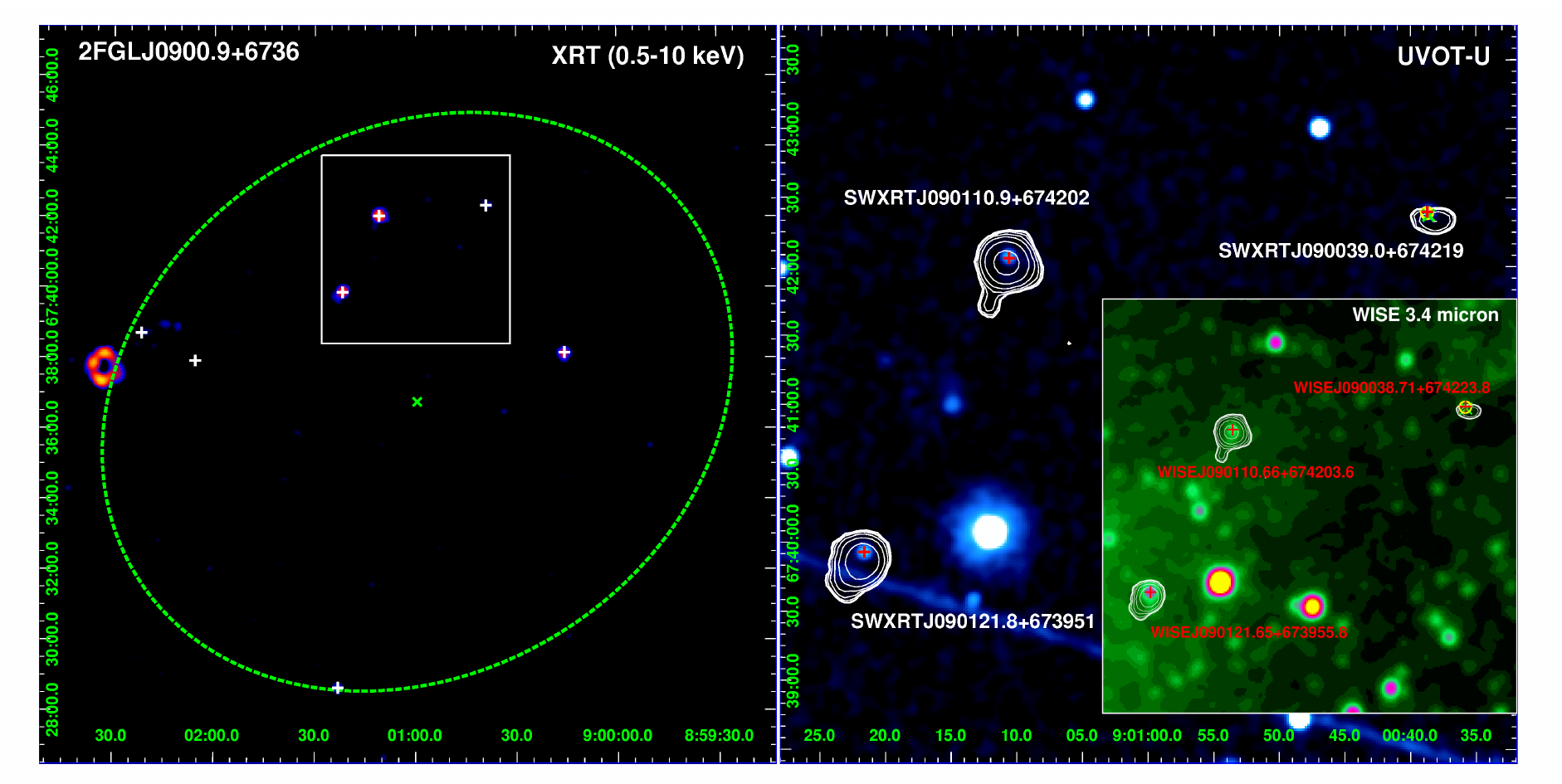}
\end{center}
\caption{(left frame) Merged XRT-PC image (0.5-10 keV) of the UGS 2FGLJ0900.9+6736. The dashed green ellipse 
indicates the the positional uncertainty region at 95\% level of confidence as reported in 2FGL catalog, and the white 
crosses indicate the detected X-ray sources. The highly piled-up source on the left is a star clearly visible in UV. (right 
frame) UVOT-U image of the region indicated in the right frame with the white box, with superimposed X-ray 
contours in white. Red crosses represent WISE counterparts to X-ray sources, yellow circles represent NVSS 
counterparts and green x-crosses represent WENSS counterparts. In the inset we show the 3.4 \(\mu\)m WISE image 
of the same region of right frame, indicating in red the name of the WISE counterparts to X-ray sources.}\label{image}
\end{figure*}

Recently, \citet{2013ApJS..206...12D} proposed a classification method to identify \(\gamma\)-ray blazar candidates 
on the basis of their positions in the three-dimensional WISE color space. As a matter of fact, blazars - whose 
emission is dominated by beamed, non thermal emission - occupy a defined region in such a space, well separated 
from that occupied by other sources in which thermal emission prevails \citep{paper2,paper3}. This method, however 
can only be applied to WISE sources detected in all 4 WISE bands, i.e., 3.4, 4.6, 12 and 22 \(\mu\)m. 

Since 414 out of 610 blazars used by \citet{2013ApJS..206...12D} are 
detected in X-rays, we here use the XRT detection as additional information 
and consider the 148 sources in our catalog with WISE counterparts detected 
only in the first 3 WISE, bands; we present their projection on the 
two-dimensional [3.4]-[4.6] [4.6]-[12] color-color plane in Figure 
\ref{strip}. In order to select \(\gamma\)-ray blazar-like 
{candidate} counterparts among these sources, we evaluate the 
two-dimensional densities of known \(\gamma\)-ray {blazars} using 
the KDE technique \citep[see, e.g.,][and reference 
therein]{2004ApJS..155..257R,2009MNRAS.396..223D,2011MNRAS.418.2165L}, and 
conservatively consider as \(\gamma\)-ray blazar-like {candidate} 
counterparts those sources with WISE colors compatible with the 90\% KDE 
density contour normalized to the peak density. On the same figure we 
indicatively show the isodensity contours of generic WISE sources, clearly 
showing that \(\gamma\)-ray blazars are well separated on this color-color 
plane from others sources \citep[see also][]{paper2,paper3}.

In this way we select 64 blazar-like {candidate} counterparts lying 
in the uncertainty region of 33 UGSs. In particular, among these 33 UGSs 
the sources 2FGLJ0200.4-4105, 2FGLJ1033.5-5032 2FGLJ1328.5-4728, 
2FGLJ1738.9+8716, 2FGLJ2228.6-1633 and 2FGLJ2246.3+1549 feature a unique 
X-ray counterpart, and are therefore considered \(\gamma\)-ray blazar 
candidates.

We note that \citet{2013arXiv1303.3585M} applied the classification method 
proposed by \citet{2013ApJS..206...12D} to the same UGSs sample discussed 
here, selecting 75 blazar-like WISE sources (see Sect. \ref{sec:blazar}). 
Among these 75 sources 28 have an X-ray counterpart in our catalog, and 26 
out of these 28 - with the exceptions of SWXRTJ011619.2-615344 and 
SWXRTJ174507.7+015442 - are also selected as \(\gamma\)-ray blazar-like  
{candidate} counterparts with the KDE technique proposed here. This 
is an excellent agreement, considering that the method proposed by 
\citet{2013ApJS..206...12D} makes use of a three-dimensional modelization 
in the \textit{Principal Component} space, while the KDE contours in Figure 
\ref{strip} represent a two-dimensional source density in the 
\textit{color} space \citep{paper3}. In addition, with the KDE technique we 
also select the source SWXRTJ060102.8+383829, whose radio counterpart 
WN0557.5+3838 has been classified as \(\gamma\)-ray blazar-like source by 
\citet{massaro2013b} on the basis of its low-frequency radio properties 
(see Sect. \ref{sec:blazar}). We so select 37 new \(\gamma\)-ray 
blazar-like {candidate} counterparts, marked in Table 
\ref{table_xrt} (column 10) with the ``KDE" string, and present their 
{SEDs} in Appendix \ref{app:seds}.

\section{Comparison with previous analyses}
\label{sec:comparison}

\subsection{Gamma-ray blazar candidates}
\label{sec:blazar}

As anticipated in Sect. \ref{sec:kde}, we compare our results with those of 
\citet{2013arXiv1303.3585M}, that applied the classification method 
proposed by \citet{2013ApJS..206...12D} to the same UGSs sample considered 
in this work, finding 75 blazar-like WISE {candidate} counterparts 
in the \textit{Fermi} LAT positional uncertainty region of 61 UGSs. Among 
these UGSs, for the 35 for which we have available XRT-PC observations we 
find no X-ray counterparts only for 5 of them. For the other 30 UGSs, 
\citet{2013arXiv1303.3585M} find a total of 44 blazar-like WISE 
{candidate} counterparts, and in our catalog we find X-ray 
counterparts to 28 of the latter. These sources are marked in Table 
\ref{table_xrt} (column 10) with the ``WISE" string, and their SEDs are 
presented in Appendix \ref{app:seds}. In particular, among these 30 UGSs 
the sources 2FGLJ0116.6-6153, 2FGLJ0227.7+2249, 2FGLJ0316.1-6434, 
2FGLJ0414.9-0855, 2FGLJ0723.9+2901, 2FGLJ1029.5-2022, 2FGLJ1254.2-2203, 
2FGLJ1614.8+4703, 2FGLJ1622.8-0314 and 2FGLJ1924.9-1036 feature a unique 
X-ray counterpart, and are therefore considered \(\gamma\)-ray blazar 
candidates.

We also compare our results with those of \citet{massaro2013b}, that 
investigate the low-frequency radio properties of blazars and searched for 
sources with similar radio properties combining the information derived 
from the WENSS and NVSS surveys, identifying 26 \(\gamma\)-ray blazar-like 
sources in the \textit{Fermi} LAT positional uncertainty regions of 21 
UGSs. Among these 21 objects, we have available XRT-PC observations for 17 
UGSs, and we find no X-ray sources for 3 of them. For the remaining 18 UGSs 
\citet{2013arXiv1303.3585M} find a total of 20 \(\gamma\)-ray blazar-like 
sources, and in our catalog we find {an X-ray counterpart} to 1 of 
them - WN0557.5+3838 - namely the source SWXRTJ060102.8+383829 
(NVSSJ060102+383828). This sources is marked in Table \ref{table_xrt} 
(column 10) with the ``WENSS" string, and its SED is presented in Appendix 
\ref{app:seds}. We note that SWXRTJ060102.8+383829 is the only X-ray source 
lying in the uncertainty region of the UGS 2FGLJ0600.9+3839, which we 
therefore consider a \(\gamma\)-ray blazar candidate.

We stress that these three methods to identify \(\gamma\)-ray blazar-like 
sources - namely, the one proposed by \citet{2013ApJS..206...12D} based on 
three-dimensional WISE colors space, the one proposed by 
\citet{massaro2013b} based on low-frequency radio properties, and the KDE 
technique applied to the two-dimensional WISE colors space - do not 
necessarily select the same sources (see Tables \ref{radio_blazar_table} 
and \ref{radio_blazar_sas_table}), nor do they necessarily select the 
brighter X-ray {candidate} counterpart of the UGS. As an example we 
show in the left frame of Figure \ref{image} the merged XRT-PC image 
(0.5-10 keV) of the UGS 2FGLJ0900.9+6736 (the bright, highly piled-up 
source on the left is a star, clearly visible in UV). The dashed green  
ellipse indicates the positional uncertainty region at 95\% level of 
confidence as reported in 2FGL catalog. In the right frame of the same 
Figure we show the UVOT-U merged image of the region indicated in the left 
frame with the white box, with superimposed X-ray contours. This region 
contains the \(\gamma\)-ray blazar-like source SWXRTJ090121.8+673951, with 
a count rate of \(5.87\pm 1.10\,{10}^{-3}\mbox{ ph}\mbox{ s}^{-1}\), 
selected on the basis of the IR colors of its WISE counterpart. However, 
the brighter X-ray source detected in the LAT positional uncertainty region 
is SWXRTJ090110.9+674202, with a count rate of \(7.07\pm 1.10 \, {10}^{-3} 
\mbox{ ph} \mbox{ s}^{-1}\) is not selected as \(\gamma\)-ray blazar-like 
source, as well as SWXRTJ090039.0+674219, with a count rate of 
\(1.52\pm 0.53\,{10}^{-3}\mbox{ ph}\mbox{ s}^{-1}\), which is the only 
X-ray source in the LAT positional uncertainty region that shows a radio 
counterpart within the XTR-PC positional error - namely NVSSJ090038+674223 
(indicated with a yellow circle) and WN0856.1+6754 (indicated with a green 
x-cross). Finally, SWXRTJ090123.0+672838 (the southernmost X-ray source 
shown in the left frame of \ref{image}, is selected as a \(\gamma\)-ray 
blazar-like source with the KDE technique and has a count rate of \(1.33\pm 
0.49\,{10}^{-3}\mbox{ ph}\mbox{ s}^{-1}\).

\subsection{Sources without counterparts}
\label{sec:nosources}

\begin{figure}
\begin{center}
\includegraphics[scale=0.3]{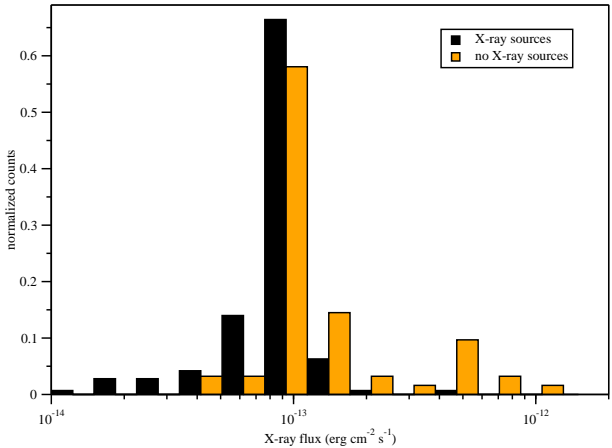}
\end{center}
\caption{X-ray fluxes reached by XRT-PC observation of the 62 UGSs that 
show no X-ray counterpart falling in the \textit{Fermi} LAT 
{positional} uncertainty region (orange bars) compared with X-ray 
fluxes reached in the 143 UGSs that show at least {one} X-ray 
{candidate} counterpart (black bars). The flux limit is estimated 
with the same spectral model considered in Sect. \ref{sec:ugs} (see Figure 
\ref{exposures_vs_fluxes}).}\label{exposures_noobs}
\end{figure}

As anticipated in Section \ref{sec:ugs}, 62 UGSs of our sample (most of 
them lying on the Galactic plane), although featuring XRT-PC observations, 
show no X-ray counterpart. The X-ray fluxes reached by XRT-PC observations 
of these sources are presented in Figure \ref{exposures_noobs} in 
comparison with the X-ray fluxes reached for UGSs that show X-ray 
{candidate} counterparts. The flux limit is estimated with the same 
spectral model considered in Sect. \ref{sec:ugs} (see Figure 
\ref{exposures_vs_fluxes}). We see that the observations of sources that 
show at least one X-ray {candidate} counterpart reach lower fluxes 
\(\sim {10}^{-14}\mbox{ erg}\mbox{ cm}^{-2}\mbox{ s}^{-1}\) with respect to 
observations of sources that show no X-ray counterparts, the latter 
reaching fluxes \(\sim 4\times{10}^{-14}\mbox{ erg}\mbox{ cm}^{-2}\mbox{ 
s}^{-1}\). The two observations, however peak at the same X-ray flux of 
\(\sim{10}^{-13}\mbox{ erg}\mbox{ cm}^{-2}\mbox{ s}^{-1}\). In 
particular we have 45 UGSs that, despite a total exposure time \(>3\mbox{ 
ks}\), do not show any X-ray counterpart. Moreover, we note that 7 of these 
UGSs - namely 2FGLJ0002.7+6220, 2FGLJ0248.5+5131, 2FGLJ0332.1+6309, 
2FGLJ0431.5+3622, 2FGLJ0602.7-4011, 2FGLJ1032.9-8401 and 2FGLJ1759.2-3853 - 
have a \(\gamma\)-ray blazar-like {candidate} counterpart in their 
positional uncertainty region, as reported by \citet{2013arXiv1303.3585M} 
and \citet{massaro2013b}.

Moreover, we have 35 UGSs that, in {their} \textit{Fermi} LAT 
{positional} uncertainty region, show X-ray {candidate} 
counterparts in XRT-PC observations, but without lower energy counterparts 
in either UVOT observations or the catalogs we described in Section 
\ref{sec:ugs}. To take into account the astrometric uncertainties of these 
catalogs, we searched for counterpart of these sources using a searching 
radius equal to three times the positional error obtained with 
\textsc{xrtcentroid}, yielding 6 UGS - namely 2FGLJ0239.5+1324, 
2FGLJ0644.6+6034, 2FGLJ0745.5+7910, 2FGLJ1544.5-1126, 2FGLJ1842.3-5839 and 
2FGLJ2133.5-6431 - that show an X-ray {candidate} counterpart 
without lower energy counterparts.We present a list of these sources in 
Table \ref{pulsar_table}, that can be useful for follow up observations 
aiming at determine their exact nature.

\subsection{Comparison with 1FGL catalog}
\label{sec:1fgl}

We note that among the 299 UGSs analyzed, there are 66 sources that were 
also unidentified according to the investigation performed in the first 
\textit{Fermi} \(\gamma\)-ray catalog (1FGL) but have been classified as 
active galactic nuclei (AGNs) or as pulsars (PSRs) using two different 
statistical approaches: the Classification Tree and the Logistic regression 
analyses \citep[see][and references therein]{2012ApJ...753...83A}. In 
particular, 38 out of the 66 show \(\gamma\)-ray properties similar to 
those of others \(\gamma\)-ray AGNs while 11 are potential PSRs with the 
remaining 17 of unknown origin.  

For the 49 UGSs classified on the basis of the above statistical methods, 
we performed a comparison with our results in particular to check if the 
2FGL sources having in their uncertainty region {an} X-ray source 
whose IR counterpart {features} blazar-like WISE colors according to 
the KDE technique illustrated in Sect. \ref{sec:kde} were also classified 
as AGNs according to the results of \citet{2012ApJ...753...83A}. We found 
that 8 out of 33 UGSs we associate with a \(\gamma\)-ray blazar-like 
{source} are also classified as AGNs, all of them with a probability 
systematically higher than 60\%. There is only one case (i.e., 2FGL 
1328.5-4728) in which the statistical procedures assigned a PSR 
classification, with a low probability (i.e., 53\%) while the KDE method 
identified the X-ray {candidate} counterpart of the \textit{Fermi} 
source as {a} blazar-like object.

\section{Summary and conclusions}
\label{sec:summary}

In this work we present a catalog of {X-ray sources lying in the 
positional uncertainty regions of} the 299 UGSs reported in the 2FGL 
catalog without any \(\gamma\)-ray analysis flag. To this end, we made use 
of all available observations performed by \textit{Swift} XRT 
in PC mode up to March 31, 2013, that where available for 205 UGSs. In 
order to detect the fainter sources, we merged all the observations 
corresponding to each UGSs, and applied to these merged observations 
different detection algorithms (i.e., \textsc{ximage} \textsc{detect} and 
\textsc{sosta}). The source list was cleaned from spurious and extended 
sources by visual inspection of all the observations, to yield a final 
catalog of 357 X-ray sources. We searched several major radio, IR, optical 
and UV surveys for any possible counterparts within the positional error of 
our X-ray sources to obtain additional information on their nature, 
providing a comprehensive list of X-ray sources with multi-wavelength 
properties.

The main results of our analysis can be summarized as follows:

\begin{itemize}
\item We find X-ray {candidate} counterparts for \(\sim 70\)\% of 
the UGSs investigated. In particular, we have \(\sim 25\)\% UGSs featuring 
a single X-ray counterpart and \(\sim 45\)\% UGSs featuring multiple X-ray 
{candidate} counterparts falling in the positional uncertainty 
region at 95\% level of confidence.
\item For each X-ray source we also looked in the corresponding UVOT merged 
images for UV-optical counterparts performing sources photometry, and 
finding UV-optical counterparts to \(\sim 71\)\% of the X-ray sources in 
our catalog.
\item We find no X-ray counterparts for 62 UGSs in our sample (\(\sim 
30\)\%), 46 of which have a total exposure \(\geq 3\mbox{ ks}\).
\item Comparing our results with \citet{2013arXiv1303.3585M} and 
\citet{massaro2013b} we find X-ray {candidate} counterparts to 29 
sources classified as \(\gamma\)-ray blazar-like.
\item Applying the KDE technique to IR colors of WISE counterparts, we 
obtain an additional list of 37 \(\gamma\)-ray blazar-like sources for 33 
UGSs (29 with a unique candidate and 4 with a double candidate). In 
particular, 10 out of these 33 2FGL sources have radio counterparts, and 
for 4 UGSs out of 33 we add a different \(\gamma\)-ray blazar-like sources 
from those selected by \citet{2013arXiv1303.3585M} and \citet{massaro2013b}.
\item Among the 51 UGSs that have a single X-ray {counterpart}, 17 
have their X-ray counterpart selected as \(\gamma\)-ray blazar-like source
with the three methods discussed above, and are there considered as 
\(\gamma\)-ray blazar candidates.
\item The source 2FGL1328.5-4728, a \(\gamma\)-ray blazar candidate 
selected with the KDE technique, is classified as PSR by 
\citet{2012ApJ...753...83A}.
\end{itemize}

Even though blazars are expected to be bright in X-rays, the methods 
discussed here to find \(\gamma\)-ray blazar-like sources in UGSs 
uncertainty regions show that this is not always the case.

We note that 39 2FGL sources in our sample are in common with the analysis 
of 1FLG UGSs by \citep{2013arXiv1307.5581T}. Comparing our results with 
\citet{2012ApJ...753...83A} we note that 38 2FGL sources in our sample are 
classified as AGN the 1FGL catalog with high level of confidence, 11 2FGL 
sources in our sample are classified as PSR with low level of confidence, 
and 17 2FGL sources in our sample are unclassified. In particular, 8 2FGL 
sources with a \(\gamma\)-ray blazar-like source selected with the KDE 
technique are classified as AGN by \citet{2012ApJ...753...83A}.

Ground-based, optical and near IR, spectroscopic follow up observations 
will be planned for the \textit{Swift} XRT sources selected as 
\(\gamma\)-ray blazar-like {candidate}  counterparts because they 
are crucial to confirm the nature of the selected sources and to obtain 
their redshift, as shown for the unidentified INTEGRAL and \textit{Swift} 
sources \citep[e.g.,][and references therein]{2012A&A...538A.123M,2012A&A...545A.101P}.

~\\
\acknowledgements
We acknowledge useful comments and suggestions by our anonymous referee.
The authors gratefully acknowledge A. Falcone for the \textit{Swift} XRT Survey of \textit{Fermi} Unassociated Sources that produced most of the observations used in this work. 
F. Massaro is grateful to M. Ajello for his support.
The work is supported by the NASA grants NNX12AO97G.
R. D'Abrusco gratefully acknowledges the financial 
support of the US Virtual Astronomical Observatory, which is sponsored by the
National Science Foundation and the National Aeronautics and Space Administration.
H. A. Smith acknowledges partial support from NASA/JPL grant RSA 1369566.
The work by G. Tosti is supported by the ASI/INAF contract I/005/12/0.
TOPCAT\footnote{\href{http://www.star.bris.ac.uk/~mbt/topcat/}
{http://www.star.bris.ac.uk/~mbt/topcat/}} \citep{2005ASPC..347...29T} for the preparation and manipulation of the tabular data and the images.
The WENSS project was a collaboration between the Netherlands Foundation 
for Research in Astronomy and the Leiden Observatory. 
We acknowledge the WENSS team consisted of Ger de Bruyn, Yuan Tang, 
Roeland Rengelink, George Miley, Huub Rottgering, Malcolm Bremer, 
Martin Bremer, Wim Brouw, Ernst Raimond and David Fullagar 
for the extensive work aimed at producing the WENSS catalog.
Part of this work is based on archival data, software or on-line services provided by the ASI Science Data Center.
This research has made use of data obtained from the High Energy Astrophysics Science Archive
Research Center (HEASARC) provided by NASA's Goddard
Space Flight Center; the SIMBAD database operated at CDS,
Strasbourg, France; the NASA/IPAC Extragalactic Database
(NED) operated by the Jet Propulsion Laboratory, California
Institute of Technology, under contract with the National Aeronautics and Space Administration.
This research has made use of software provided by the Chandra X-ray Center (CXC) in the application packages CIAO, ChIPS, and Sherpa.
Part of this work is based on the NVSS (NRAO VLA Sky Survey);
The National Radio Astronomy Observatory is operated by Associated Universities,
Inc., under contract with the National Science Foundation. 
This publication makes use of data products from the Two Micron All Sky Survey, which is a joint project of the University of 
Massachusetts and the Infrared Processing and Analysis Center/California Institute of Technology, funded by the National Aeronautics 
and Space Administration and the National Science Foundation.
This publication makes use of data products from the Wide-field Infrared Survey Explorer, 
which is a joint project of the University of California, Los Angeles, and 
the Jet Propulsion Laboratory/California Institute of Technology, 
funded by the National Aeronautics and Space Administration.
Funding for the SDSS and SDSS-II has been provided by the Alfred P. Sloan Foundation, 
the Participating Institutions, the National Science Foundation, the U.S. Department of Energy, 
the National Aeronautics and Space Administration, the Japanese Monbukagakusho, 
the Max Planck Society, and the Higher Education Funding Council for England. 
The SDSS Web Site is http://www.sdss.org/.
The SDSS is managed by the Astrophysical Research Consortium for the Participating Institutions. 
The Participating Institutions are the American Museum of Natural History, 
Astrophysical Institute Potsdam, University of Basel, University of Cambridge, 
Case Western Reserve University, University of Chicago, Drexel University, 
Fermilab, the Institute for Advanced Study, the Japan Participation Group, 
Johns Hopkins University, the Joint Institute for Nuclear Astrophysics, 
the Kavli Institute for Particle Astrophysics and Cosmology, the Korean Scientist Group, 
the Chinese Academy of Sciences (LAMOST), Los Alamos National Laboratory, 
the Max-Planck-Institute for Astronomy (MPIA), the Max-Planck-Institute for Astrophysics (MPA), 
New Mexico State University, Ohio State University, University of Pittsburgh, 
University of Portsmouth, Princeton University, the United States Naval Observatory, 
and the University of Washington.
The United Kingdom Infrared Telescope is operated by the Joint Astronomy Centre on behalf of the Science and Technology Facilities Council of the U.K.
The CSS survey is funded by the National Aeronautics and Space
Administration under Grant No. NNG05GF22G issued through the Science
Mission Directorate Near-Earth Objects Observations Program.  The CRTS
survey is supported by the U.S.~National Science Foundation under
grants AST-0909182.

{}

\newpage

\begin{appendix}

\section{Catalog tables}\label{app:tables}

Here we present the catalog of X-ray sources with their main properties.

In Table \ref{table_xrt} we list all the X-ray sources found in XRT-PC observations in the positional uncertainty region 
of each UGS. The columns contain the following information: (1) NAME XRT: source designation as described in 
Section \ref{sec:swift} and corresponding 2FGL UGS; (2) OTHER NAME: name of the counterpart found in the catalogs 
described in Section \ref{sec:ugs}. If more than one counterpart is found, the order we choose for the alternate name 
is the following: NVSS, FIRST, SUMSS, WENSS, WISE, SDSS, 6DFGS, NED; (3) RA: right ascension as given by 
\textsc{xrtcentroid}; (4) DEC: declination as given by \textsc{xrtcentroid}; (5) ERR: positional error in arcseconds as 
given by \textsc{xrtcentroid}; (6) EXP: XRT-PC total exposure in seconds; (7) COUNTRATE: countrate and relative error 
as given by \textsc{detect} in \({10}^{-3}\mbox{ph }\mbox{s}^{-1}\); (8) SIGN: signal to noise threshold above which 
the source is detected by \textsc{detect}; (9) SOSTA: countrate and relative error as given by \textsc{sosta} in \({10}
^{-3}\mbox{ph }\mbox{s}^{-1}\); (10) SNR: signal to noise ratio as given by \textsc{sosta}; (11) NOTES: results of the 
cross-matching with the catalogs discussed in Section \ref{sec:ugs} within the positional error reported in column 
ERR: NVSS=N, FIRST=F, SUMSS=S, WENSS=W, WISE=w, 2MASS=M, UKIDSS=UK, SDSS=s, 6=6DFGS, GALEX=g, 
UVOT(optical filter)=o, UVOT(UV filter)=u, USNO-B=U, CSC=C; (12) CAND: \(\gamma\)-ray blazar-like sources according to 
\citet{2013arXiv1303.3585M} (WISE), to \citet{massaro2013b} (WENSS) and to the KDE technique as discussed in 
Section \ref{sec:blazar}; (12) REDSHIFT: redshift for the source counterpart as reported by SDSS, 6DFGS or NED.

In Table \ref{table_uvot} we list, for each source in Table \ref{table_xrt}, the properties of the UV-optical counterpart 
found in merged UVOT observations. The columns contain the following information: (1) NAME XRT: source 
designation as described in Section \ref{sec:swift} and corresponding 2FGL UGS; (2) RA: right ascension of the UVOT 
counterpart; (3) DEC: declination UVOT counterpart; (4) SEP: angular separation in arcseconds between the XRT-PC 
source and the UVOT counterpart; (5) E(B-V): galactic extinction value as derived by the Infrared Science Archive
\footnote{\href{http://irsa.ipac.caltech.edu/applications/DUST/}{http://irsa.ipac.caltech.edu/applications/DUST/}} 
(IRSA); (6) EXPV: exposure of the UVOT-V filter merged observation in seconds; (7) MAGV: UVOT-V filter magnitude 
(Vega system) and relative error as given by \textsc{uvotsource} (not corrected by galactic extinction). Upper limits are 
indicated with 0.00 errors, while * indicate filter saturation; (8) MAGVS: UVOT-V filter magnitude (Vega system) and 
relative error as given by \textsc{uvotdetect} (not corrected by galactic extinction). Upper limits are indicated with 
0.00 errors, while * indicate filter saturation; (9) EXPB, (10) MAGB, (11) MAGBS: same as columns (6), (7) and (8) but for 
UVOT-B filter; (12) EXPU, (13) MAGU, (14) MAGUS: same as columns (6), (7) and (8) but for UVOT-U filter; (15) EXPW1, 
(16) MAGW1, (17) MAGW1S: same as columns (6), (7) and (8) but for UVOT-W1 filter; (18) EXPM2, (19) MAGM2, (20) 
MAGM2S: same as columns (6), (7) and (8) but for UVOT-M2 filter; (21) EXPW2, (22) MAGW2, (23) MAGW2S: same as 
columns (6), (7) and (8) but for UVOT-W2 filter.

In Table \ref{usnomag} we list all the XRT-PC sources that features a USNO-B counterpart within the positional error 
and present the magnitudes of this counterpart. The columns contain the following information: (1) NAME XRT: source 
designation as described in Section \ref{sec:swift}; (2) B1: first epoch blue magnitude; (3) B2: second epoch blue 
magnitude; (4) R1: first epoch red magnitude; (5) R2: second epoch red magnitude; (6) I: second epoch near-IR 
magnitude.

In Table \ref{pulsar_table} we list all UGS that, although featuring XRT-PC observations, show no X-ray counterpart.
The columns contain the following information: (1) NAME 2FGL: UGS name as reported in the 2FGL, with boldface 
indicating those sources that have a \(\gamma\)-ray blazar-like {candidate} counterpart in their positional uncertainty region 
as reported by \citet{2013arXiv1303.3585M} and \citet{massaro2013b}; (2) EXP: XRT-PC total exposure in seconds.

\begin{sidewaystable*}
\caption{Sample catalog of XRT-PC detected sources in the positional uncertainty region of each UGS as reported in the 2FGL. Column description is given in Appendix \ref{app:tables}.}\label{table_xrt}
\resizebox{\textwidth}{!}{
\begin{tabular}{llcccccccclcc}
\hline
\hline
NAME XRT & OTHER NAME & RA & DEC & ERR & EXP & COUNTRATE & SIGN & SOSTA & SNR & NOTES & CAND & REDSHIFT \\
& & J2000 & J2000 & arcsec & s & \({10}^{-3}\mbox{ph }\mbox{s}^{-1}\) & & \({10}^{-3}\mbox{ph }\mbox{s}^{-1}\) & & & & \\
\hline
\\
2FGLJ0031.0+0724 & & & & & & & & & & & \\
\hline
SWXRTJ003054.8+072324 & WISEJ003054.91+072323.7 & 00:30:54.817 & +07:23:24.33 & 5.70 & 8137 & 1.87(0.60) & 3 & 1.55(0.59) & 2.61 & w,M,UK,s,g,o,u,U & & \\
SWXRTJ003113.1+073143 & SDSSJ003112.79+073143.6 & 00:31:13.114 & +07:31:43.31 & 4.78 & 7571 & 5.95(1.00) & 5 & 6.42(1.10) & 6.07 & UK,s,o,u & & \\
SWXRTJ003119.9+072452 & FIRSTJ003119.7+072454 & 00:31:19.927 & +07:24:52.48 & 4.06 & 7914 & 16.50(1.60) & 5 & 16.19(1.60) & 9.92 & F,w,UK,s,o,u,U & KDE & \\
\hline
\\
2FGLJ0039.1+4331 & & & & & & & & & & & \\
\hline
SWXRTJ003858.3+432947 & WISEJ003858.27+432947.0 & 00:38:58.261 & +43:29:46.82 & 5.11 & 3956 & 6.98(1.60) & 4 & 7.60(1.60) & 4.61 & w,M,o,U & WISE & \\
SWXRTJ003908.5+433027 & & 00:39:08.475 & +43:30:27.46 & 5.76 & 4093 & 2.84(1.10) & 2 & 2.84(1.10) & 2.57 & & & \\
SWXRTJ003938.4+433446 & & 00:39:38.387 & +43:34:46.23 & 5.35 & 4265 & 2.35(0.94) & 2 & 2.58(1.00) & 2.55 & & & \\
\hline
\hline
\end{tabular}}
\end{sidewaystable*}

\begin{sidewaystable}
\caption{Sample of UVOT counterparts to the XRT-PC detected sources with their photometric properties. Column description is given in Appendix \ref{app:tables}.}\label{table_uvot}
\resizebox{\textwidth}{!}{
\begin{tabular}{lcccccccccccccccccccccccc}
\hline
\hline
NAME XRT & RA & DEC & SEP & E(B-V) & EXPV & MAGV & MAGVS & EXPB & MAGB & MAGBS & EXPU & MAGU & MAGUS & EXPW1 & MAGW1 & MAGW1S & EXPM2 & MAGM2 & MAGM2S & EXPW2 & MAGW2 & MAGW2S \\
& J2000 & J2000 & arcsec & mag & s & mag & mag & s & mag & mag & s & mag & mag & s & mag & mag & s & mag & mag & s & mag & mag & \\
\hline
\\
2FGLJ0031.0+0724 & & & & & & & & & & & & & & & & & & & & & & \\
\hline
SWXRTJ003054.8+072324 & 0:30:54.858 & +7:23:24.00 & 0.69 & 0.0316 & - & - & - & - & - & - & 3005 & 19.50(0.13) & 19.90(0.14) & 4537 & 20.32(0.18) & 21.02(0.20) & 470 & 20.05(0.34) & - & - & - & - \\
SWXRTJ003113.1+073143 & 0:31:12.807 & +7:31:43.22 & 4.57 & 0.0376 & - & - & - & - & - & - & 3005 & 18.90(0.08) & - & 4537 & 18.65(0.06) & - & 470 & 18.63(0.14) & 18.45(0.12) & - & - & - \\
SWXRTJ003119.9+072452 & 0:31:19.689 & +7:24:54.32 & 3.99 & 0.0354 & - & - & - & - & - & - & 3005 & 17.68(0.04) & 17.68(0.02) & 4537 & 17.83(0.04) & 17.79(0.02) & 470 & 17.98(0.10) & 17.89(0.08) & - & - & - \\
\hline
\\
2FGLJ0039.1+4331 & & & & & & & & & & & & & & & & & & & & & & \\
\hline
SWXRTJ003858.3+432947 & 0:38:58.276 & +43:29:47.19 & 0.40 & 0.0711 & - & - & - & - & - & - & 4176 & 17.82(0.04) & 17.70(0.01) & - & - & - & - & - & - & - & - & - \\
SWXRTJ003908.5+433027 & 0:39:08.475 & +43:30:27.46 & 0.00 & 0.0724 & - & - & - & - & - & - & 4176 & 21.79(0.00) & - & - & - & - & - & - & - & - & - & - \\
SWXRTJ003938.4+433446 & 0:39:38.387 & +43:34:46.23 & 0.00 & 0.0739 & - & - & - & - & - & - & 4176 & 21.74(0.00) & - & - & - & - & - & - & - & - & - & - \\
\hline
\hline
\end{tabular}}
\end{sidewaystable}

\begin{table}
\caption{Sample of XRT-PC sources feautring a USNO-B counterpart within the positional error. Column description is given in Appendix \ref{app:tables}.}\label{usnomag}
\begin{center}
\begin{tabular}[t]{lccccc}
\hline
\hline
NAME XRT & B1 & B2 & R1 & R2 & I\\
\hline
SWXRTJ003054.8+072324 & 19.85 & 18.32 & 19.48 & 18.24 & 18.28 \\
SWXRTJ003119.9+072452 & 19.03 & 18.17 & 19.84 & 18.63 & 18.67 \\
\hline
\hline
\end{tabular}
\end{center}
\end{table}

\begin{table}
\caption{UGSs without XRT-PC counterparts in the \textit{Fermi} LAT positional uncertainty region. In boldface we indicate those sources that have a \(\gamma\)-ray blazar-like {candidate} counterpart in their uncertainty region as reported by \citet{2013arXiv1303.3585M} and \citet{massaro2013b}.}\label{pulsar_table}
\begin{center}
\begin{tabular}[t]{lc}
\hline
\hline
NAME 2FGL & EXP \\
                   & s \\
\hline
\textbf{2FGLJ0002.7+6220} & 4817  \\
2FGLJ0032.7-5521 & 3966  \\
2FGLJ0106.5+4854 & 2889  \\
2FGLJ0237.9+5238 & 4445  \\
\textbf{2FGLJ0248.5+5131} & 2407  \\
2FGLJ0312.8+2013 & 4099  \\
\textbf{2FGLJ0332.1+6309} & 5150  \\
2FGLJ0340.5+5307 & 4977  \\
2FGLJ0359.5+5410 & 4320  \\
2FGLJ0418.9+6636 & 5319  \\
2FGLJ0426.7+5434 & 4380  \\
\textbf{2FGLJ0431.5+3622} & 4433  \\
2FGLJ0516.7+2634 & 4632  \\
2FGLJ0524.1+2843 & 3699  \\
2FGLJ0545.6+6018 & 3988  \\
2FGLJ0600.8-1949 &  979  \\
\textbf{2FGLJ0602.7-4011} & 1930  \\
2FGLJ0854.7-4501 & 4997  \\
2FGLJ0953.6-1504 & 3504  \\
\textbf{2FGLJ1032.9-8401} & 7999  \\
2FGLJ1208.5-6240 & 3738  \\
2FGLJ1306.2-6044 & 4867  \\
2FGLJ1400.2-2412 &  659  \\
2FGLJ1400.7-1438 &  417  \\
2FGLJ1410.4+7411 & 3611  \\
2FGLJ1422.3-6841 & 3421  \\
2FGLJ1458.5-2121 & 3374  \\
2FGLJ1513.9-2256 & 3316  \\
2FGLJ1521.8-5735 & 5434  \\
2FGLJ1601.1-4220 & 3394  \\
2FGLJ1617.5-2657 & 142 \\
2FGLJ1620.8-4928 &  577  \\
2FGLJ1624.1-4040 & 3399  \\
2FGLJ1646.7-1333 & 1508  \\
2FGLJ1702.5-5654 & 3197  \\
2FGLJ1712.4-3941 &  604  \\
2FGLJ1744.1-7620 & 4682  \\
2FGLJ1747.6+0324 & 3778  \\
2FGLJ1748.9-3923 &  574  \\
2FGLJ1757.5-6028 & 4011  \\
\textbf{2FGLJ1759.2-3853} & 192 \\
2FGLJ1808.3-3356 &  727  \\
2FGLJ1816.5+4511 & 4795  \\
2FGLJ1821.8+0830 &  380  \\
2FGLJ1849.3-0055 &  487  \\
2FGLJ1901.1+0427 & 2957  \\
2FGLJ1902.7-7053 & 3256  \\
2FGLJ1906.5+0720 & 10728 \\
2FGLJ1917.0-3027 & 3666  \\
2FGLJ1949.4-1457 & 4031  \\
2FGLJ2017.5-1618 & 3656  \\
2FGLJ2018.0+3626 & 4368  \\
2FGLJ2028.3+3332 & 10386 \\
2FGLJ2041.2+4735 & 3629  \\
2FGLJ2044.4-4757 & 3771  \\
2FGLJ2107.8+3652 & 4865  \\
2FGLJ2111.3+4605 & 5909  \\
2FGLJ2112.5-3042 & 2844  \\
2FGLJ2117.5+3730 & 3521  \\
2FGLJ2139.8+4714 & 3189  \\
2FGLJ2213.7-4754 & 3326  \\
2FGLJ2347.2+0707 & 3047  \\
\hline
\hline
\end{tabular}
\end{center}
\end{table}

\newpage

\section{Blazar-like sources spectral energy distributions}\label{app:seds}

Here we present the XRT-PC counterparts of \(\gamma\)-ray blazar-like sources, with their SEDs.

In Table \ref{radio_blazar_table} we list the 30 XRT-PC counterparts of \(\gamma\)-ray blazar-like sources according to 
the classification methods proposed by \citet{2013ApJS..206...12D} and \citet{massaro2013b}. In boldface we indicate 
sources with radio counterparts within the positional error listed in Table \ref{table_xrt}. Columns contain the 
following information: (1) NAME 2FGL: UGS name as reported in the 2FGL; (2) NAME XRT: source designation as 
described in Section \ref{sec:swift}; (3) ALT NAME: name of the WISE counterpart (as reported by WISE All-Sky data 
catalog, \citealt{cutri12b}) or of the WENSS counterpart (as reported by WENSS catalog, \citealt{1997A&AS..124..259R}) 
closer to the XRT-PC coordinates (as reported in Table \ref{table_xrt}); (4) CLASS: for \(\gamma\)-ray blazar-like sources selected by \citet{2013ApJS..206...12D}, every source is assigned to class A, B, or C depending on the 
probability of the WISE source to be compatible with the model of the WISE \textit{Fermi} Blazar (WFB) locus: class A 
sources are considered the most probable candidate blazars for the high-energy source, while class B and class C 
sources are less compatible with the WFB locus but are still deemed as candidate blazars. For \(\gamma\)-ray blazar-like sources selected by \citet{massaro2013b}, with A we indicate radio sources having \(-1.00\leq \alpha^{1400}_{325}\leq 0.55\) and with B those with \(0.55\leq \alpha^{1400}_{325}\leq 0.65\), where \(\alpha^{1400}_{325}\) is 
the radio spectral index between 325 MHz and 1.4 GHz; (4) TYPE: classification of the candidate blazar according 
to \citet{2013ApJS..206...12D} based on the compatibility of the WISE source with the regions of the WFB locus model. 
BZB and BZQ indicate the regions dominated by BL Lac objects and FSRQs sources respectively, while MIXED indicate 
the region where the population is mixed in terms of spectral classes.

In Table \ref{radio_blazar_sas_table} we list the 44 XRT-PC counterparts of \(\gamma\)-ray blazar-like sources according to thee KDE technique illustrated in Sect. \ref{sec:kde}. In boldface we indicate sources with radio 
counterparts within the positional error listed in Table \ref{table_xrt}. Columns contain the following information: (1) 
NAME 2FGL: UGS name as reported in the 2FGL; (2) NAME XRT: source designation as described in Section 
\ref{sec:swift}; (3) WISE NAME: name of the WISE counterpart (as reported by WISE All-Sky data catalog).

\begin{figure}
\caption{Sample SEDs of \(\gamma\)-ray blazar-like sources listed in Table \ref{radio_blazar_table} that have a radio counterpart within their XRT positional error. Symbol description is given in Appendix \ref{app:seds}.}\label{radio_seds}
\begin{center}
\includegraphics[scale=0.39]{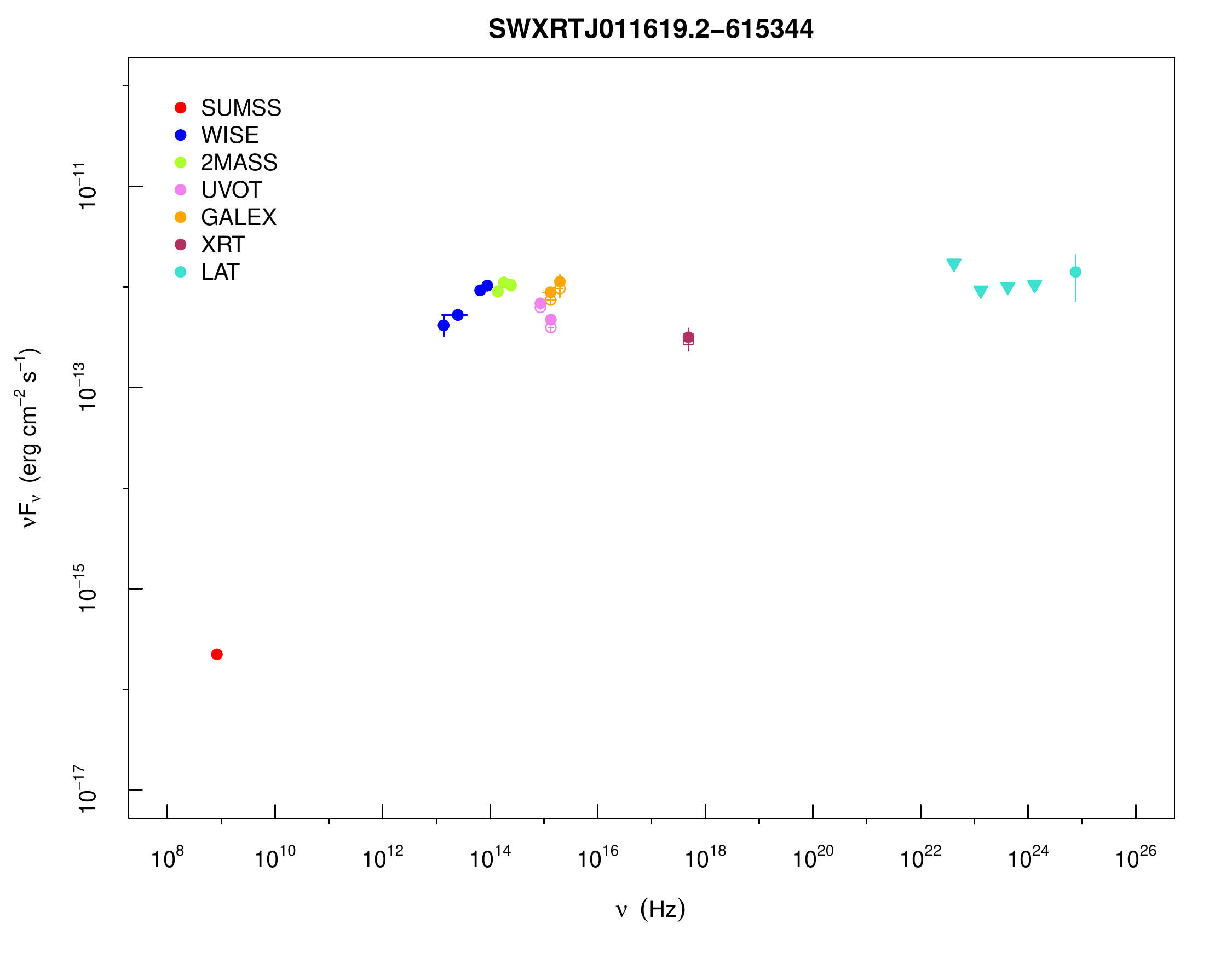}
\includegraphics[scale=0.39]{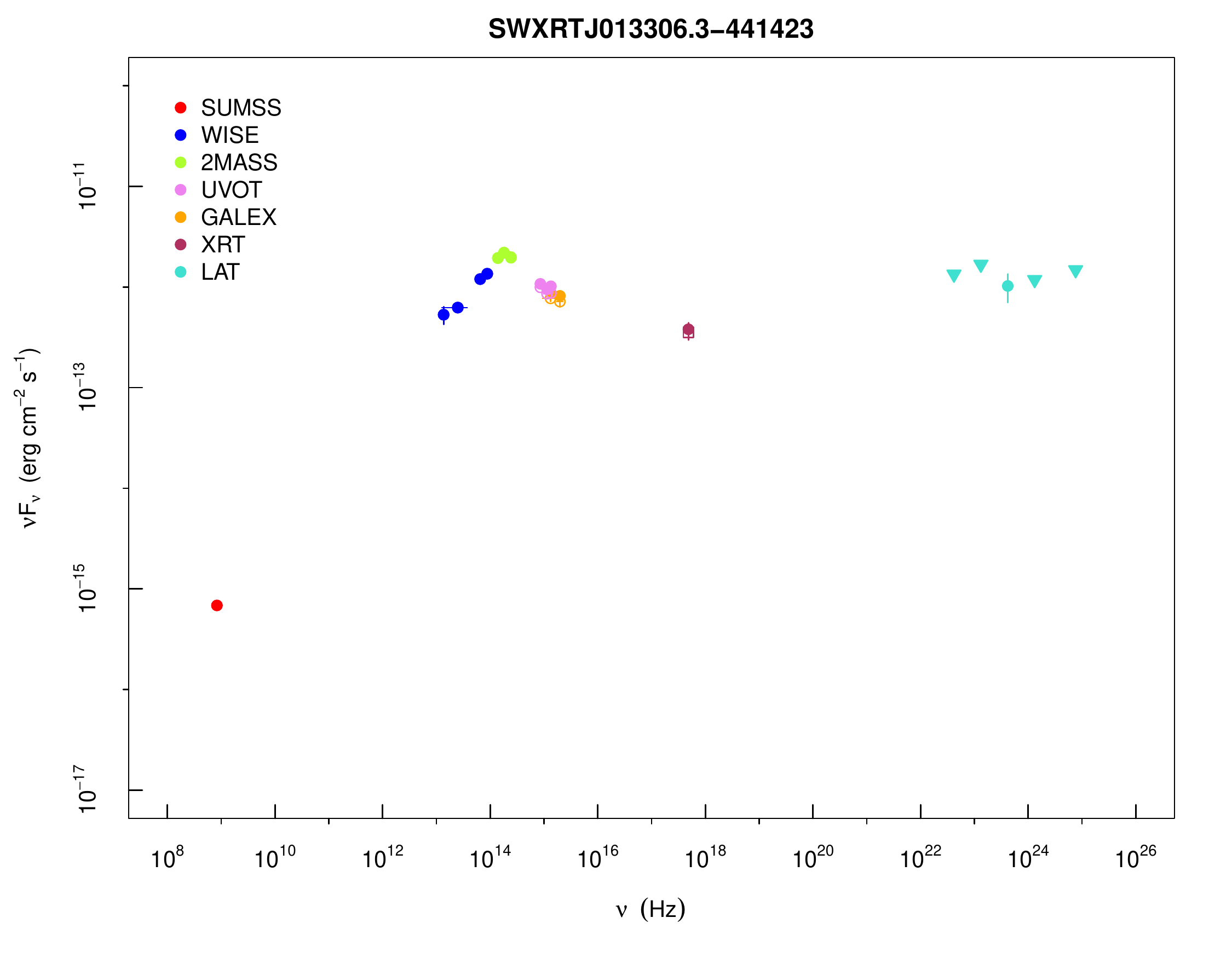}
\end{center}
\end{figure}

\begin{figure}
\caption{Sample SEDs of \(\gamma\)-ray blazar-like sources listed in Table \ref{radio_blazar_table} without a radio counterpart within their XRT positional error. Symbol description is given in Appendix \ref{app:seds}.}\label{seds}
\begin{center}
\includegraphics[scale=0.39]{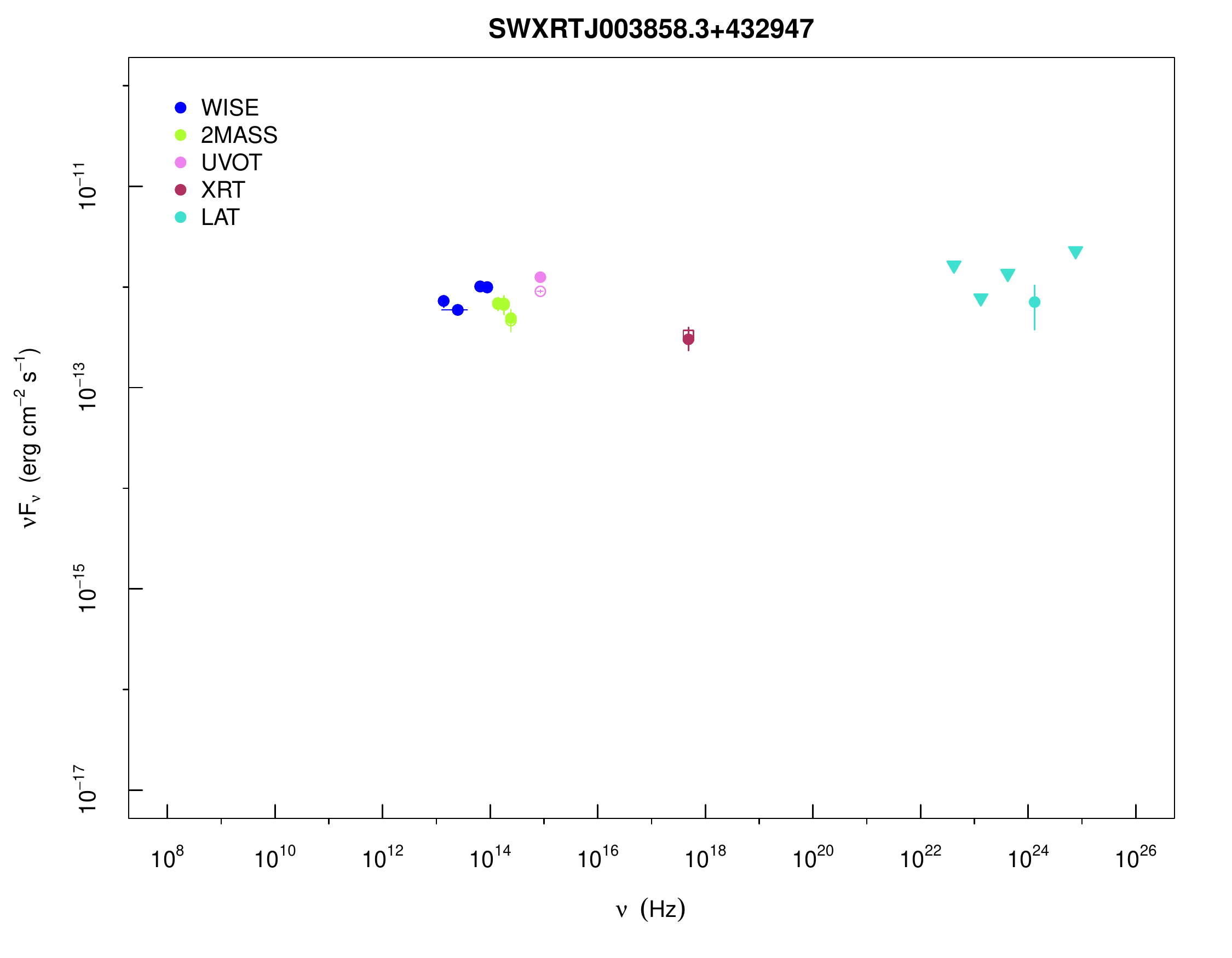}
\includegraphics[scale=0.39]{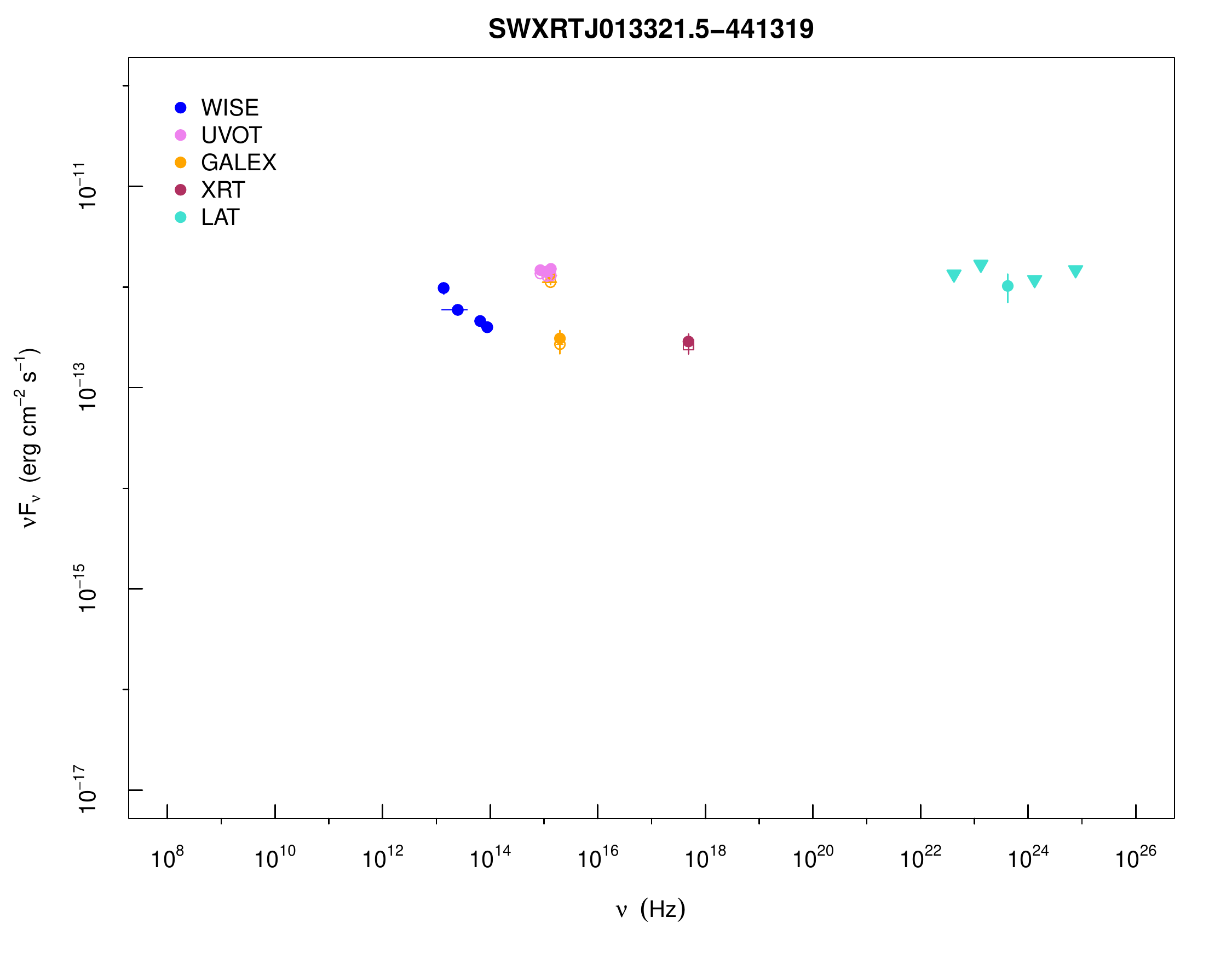}
\end{center}
\end{figure}

\begin{figure}
\caption{Sample SEDs of \(\gamma\)-ray blazar-like sources listed in Table \ref{radio_blazar_sas_table} that have a radio counterpart within their XRT positional error. Symbol description is given in Appendix \ref{app:seds}. }\label{radio_sas_seds}
\begin{center}
\includegraphics[scale=0.39]{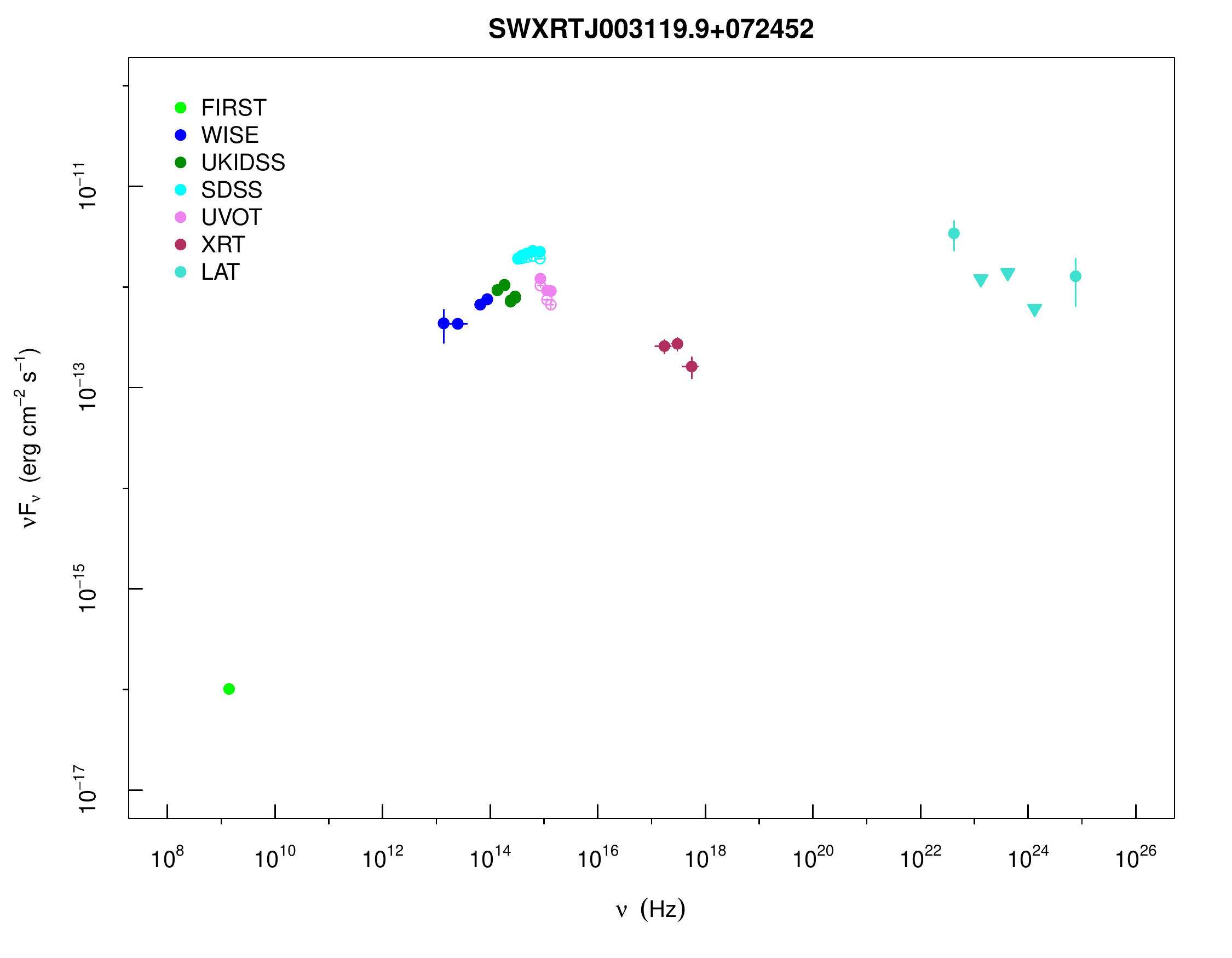}
\includegraphics[scale=0.39]{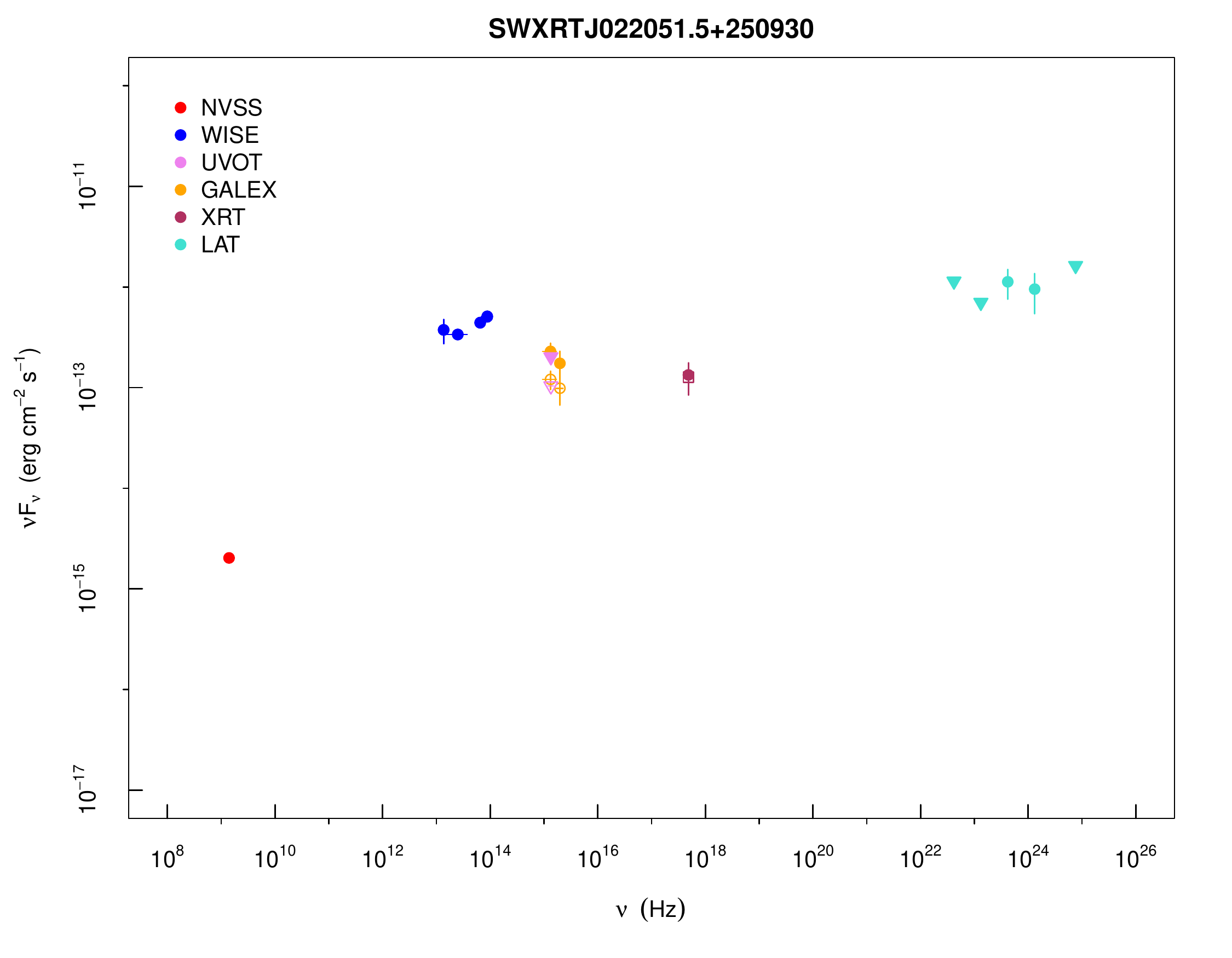}
\end{center}
\end{figure}

\begin{figure}
\caption{Sample SEDs of \(\gamma\)-ray blazar-like sources listed in Table \ref{radio_blazar_sas_table} without a radio counterpart within their XRT positional error. Symbol description is given in Appendix \ref{app:seds}. }\label{sas_seds}
\begin{center}
\includegraphics[scale=0.39]{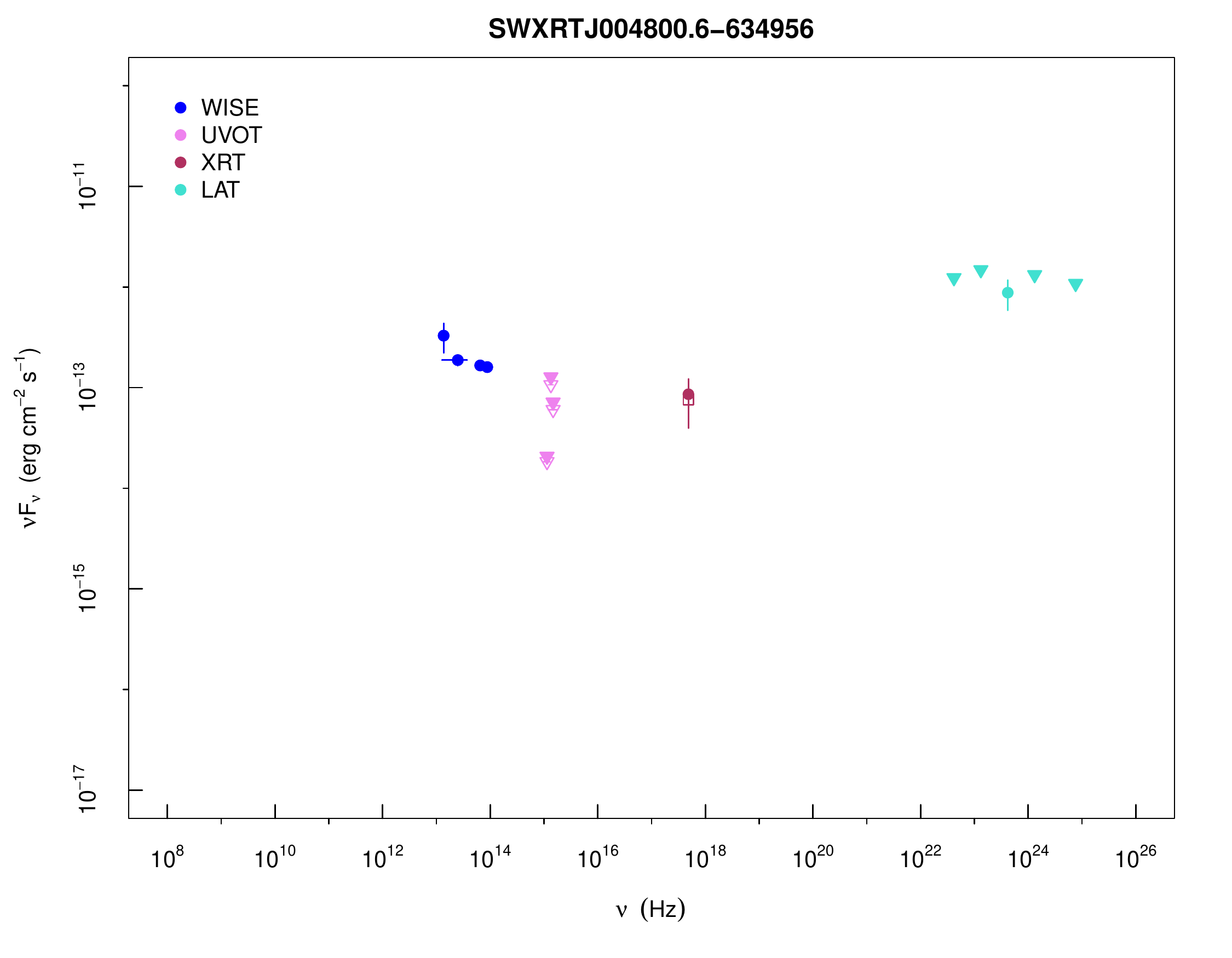}
\includegraphics[scale=0.39]{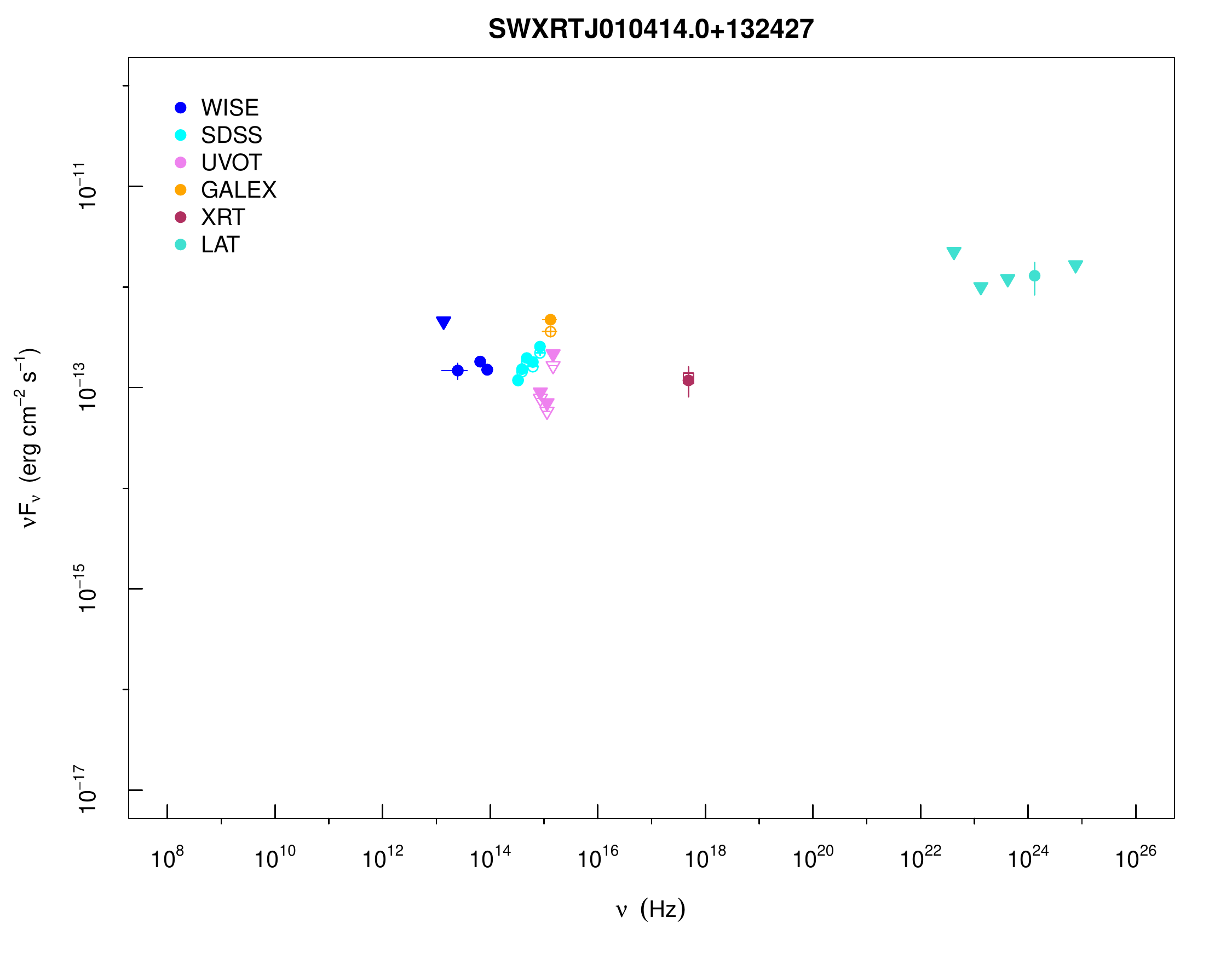}
\end{center}
\end{figure}

SEDs of the sources listed in Table \ref{radio_blazar_table} are presented in Figures \ref{radio_seds} and \ref{seds} for 
sources that feature and do not feature a radio counterpart within the XRT positional error, respectively. In the same 
way, SEDs of the sources listed in Table \ref{radio_blazar_sas_table} are presented in Figures \ref{radio_sas_seds} and 
\ref{sas_seds} for sources that feature and do not feature a radio counterpart within the XRT positional error, 
respectively. For each XRT-PC source we show the spectral points corresponding to the various counterparts we found 
in the XRT-PC positional error as reported in Table \ref{table_xrt} (see Section \ref{sec:ugs}). Circles represent 
detections, while down triangles represent upper limits, with the color code presented in the legends. For IR, optical 
and UV points we present both observed (empty symbols) and de-reddened (full symbols) fluxes, the latter obtained 
using the extinction law presented by \citet{1989ApJ...345..245C} and the galactic extinction value as derived by IRSA. 
When possible, XRT-PC spectra are obtained form events extracted with \textsc{xrtproducts} task using a 20 pixel 
radius circle centered on the coordinates reported in Table \ref{table_xrt} and background estimated from a nearby 
source-free circular region of 20 pixel radius. When the source count rate is above 0.5 counts \(\mbox{s}^{-1}\), the 
data are significantly affected by pileup in the inner part of the point-spread function \citep{2005SPIE.5898..360M}. 
To remove the pile-up contamination, we extract only events contained in an annular region centered on the source 
\citep[e.g.,][]{2007A&A...462..889P}. The inner radius of the region was determined by comparing the observed 
profiles with the analytical model derived by \citet{2005SPIE.5898..360M} and typically has a 4 or 5 pixels radius, 
while the outer radius is 20 pixels for each observation. Source spectra are binned to ensure a minimum of 20 counts 
per bin in order to ensure the validity of \(\chi^2\) statistics. We performed our spectral analysis with the 
\textsc{Sherpa}\footnote{\href{http://cxc.harvard.edu/sherpa}{http://cxc.harvard.edu/sherpa}} modeling and fitting 
application \citep{2001SPIE.4477...76F} include in the \textsc{CIAO} \citep{2006SPIE.6270E..60F} 4.5 software 
package, and with the \textsc{xspec} software package, version 12.8.0 \citep{1996ASPC..101...17A} with identical 
results. For the spectral fitting we used a model comprising an absorption component fixed to the Galactic value 
\citep{2005A&A...440..775K} and a powerlaw, and we plot intrinsic fluxes (i.e., without Galactic photoelectric 
absorption). When the extracted counts are not enough to provide acceptable spectral fits we simply converted the 
count rates reported in Table \ref{table_xrt} to 0.3-10 keV intrinsic fluxes with \textsc{PIMMS} 4.6b software, 
assuming a powerlaw spectra with spectral index 2 and an absorption component fixed to the Galactic value. In this 
case we report with a filled circle the flux corresponding to the countrate as obtained with \textsc{detect} and with an 
empty box the countrate as obtained with \textsc{sosta}.

\newpage

\begin{table}
\caption{XRT-PC counterparts to \(\gamma\)-ray blazar-like sources selected according to \citet{2013ApJS..206...12D} and \citet{massaro2013b}. In boldface we indicate sources with radio counterparts within the positional error listed in Table \ref{table_xrt}. Column description is given in Appendix \ref{app:seds}.}\label{radio_blazar_table}
\begin{center}
\begin{large}
\begin{tabular}{lllcl}
\hline
\hline
NAME 2FGL & NAME XRT & ALT NAME & CLASS & TYPE \\
\hline
2FGLJ0039.1+4331 & SWXRTJ003858.3+432947 & WISEJ003858.27+432947.0 & C & BZB \\
\hline
2FGLJ0116.6-6153 & \textbf{SWXRTJ011619.2-615344} & WISEJ011619.59-615343.5 & C & BZB \\
\hline
2FGLJ0133.4-4408 & \textbf{SWXRTJ013306.3-441423} & WISEJ013306.35-441421.3 & C & BZB \\
                              & SWXRTJ013321.5-441319 & WISEJ013321.36-441319.4 & C & BZQ \\
\hline
2FGLJ0143.6-5844 & SWXRTJ014347.1-584551 & WISEJ014347.39-584551.3 & C & BZB \\
\hline
2FGLJ0227.7+2249 & \textbf{SWXRTJ022744.0+224838} & WISEJ022744.35+224834.3 & B & BZB \\
\hline
2FGLJ0316.1-6434 & SWXRTJ031613.9-643730 & WISEJ031614.31-643731.4 & C & BZB \\
\hline
2FGLJ0409.8-0357 & \textbf{SWXRTJ040946.5-040002} & WISEJ040946.57-040003.4 & B & BZB \\
\hline
2FGLJ0414.9-0855 & SWXRTJ041457.1-085654 & WISEJ041457.01-085652.0 & C & MIXED \\
\hline
2FGLJ0600.9+3839 & \textbf{SWXRTJ060102.8+383829} & WN0557.5+3838 & B &\\
\hline
2FGLJ0644.6+6034 & SWXRTJ064459.9+603132 & WISEJ064459.38+603131.7 & C & MIXED \\
\hline
2FGLJ0723.9+2901 & \textbf{SWXRTJ072355.1+285926} & WISEJ072354.83+285929.9 & C & BZQ \\
\hline
2FGLJ0746.0-0222 & \textbf{SWXRTJ074627.1-022551} & WISEJ074627.03-022549.3 & C & BZB \\
\hline
2FGLJ0756.3-6433 & SWXRTJ075624.1-643031 & WISEJ075624.60-643030.6 & C & BZB \\
\hline
2FGLJ0838.8-2828 & SWXRTJ083842.4-282831 & WISEJ083842.77-282830.9 & C & MIXED \\
\hline
2FGLJ0900.9+6736 & SWXRTJ090121.8+673951 & WISEJ090121.65+673955.8 & C & MIXED \\
\hline
2FGLJ1013.6+3434 & SWXRTJ101256.7+343646 & WISEJ101256.54+343648.8 & C & BZB \\
\hline
2FGLJ1029.5-2022 & SWXRTJ102946.9-201808 & WISEJ102946.66-201812.6 & C & BZQ \\
\hline
2FGLJ1038.2-2423 & SWXRTJ103755.0-242543 & WISEJ103754.92-242544.5 & C & BZQ \\
\hline
2FGLJ1254.2-2203 & \textbf{SWXRTJ125422.8-220414} & WISEJ125422.47-220413.6 & C & BZB \\
\hline
2FGLJ1347.0-2956 & SWXRTJ134707.1-295844 & WISEJ134706.89-295842.3 & C & BZB \\
\hline
2FGLJ1614.8+4703 & \textbf{SWXRTJ161541.3+471110} & WISEJ161541.22+471111.8 & C & BZB \\
\hline
2FGLJ1622.8-0314 & SWXRTJ162225.3-031439 & WISEJ162225.35-031439.6 & C & BZQ \\
\hline
2FGLJ1627.8+3219 & SWXRTJ162800.3+322413 & WISEJ162800.40+322414.0 & C & BZQ \\
\hline
2FGLJ1745.6+0203 & \textbf{SWXRTJ174507.7+015442} & WISEJ174507.82+015442.5 & A & BZQ \\
                              & SWXRTJ174526.8+020532 & WISEJ174526.95+020532.7 & B & BZB \\
\hline
2FGLJ1904.8-0705 & SWXRTJ190444.6-070738 & WISEJ190444.57-070740.1 & C & MIXED \\
\hline
2FGLJ1924.9-1036 & SWXRTJ192501.8-104316 & WISEJ192501.63-104316.3 & C & BZQ \\
\hline
2FGLJ2021.5+0632 & \textbf{SWXRTJ202155.7+062913} & WISEJ202155.45+062913.7 & C & BZB \\
\hline
\hline
\end{tabular}
\end{large}
\end{center}
\end{table}

\begin{table}[h!]
\caption{XRT-PC counterparts to \(\gamma\)-ray blazar-like sources selected with KDE technique. In boldface we indicate sources with radio counterparts within the positional error listed in Table \ref{table_xrt}. Column description is given in Appendix \ref{app:seds}.}\label{radio_blazar_sas_table}
\begin{center}
\begin{large}
\begin{tabular}{lll}
\hline
\hline
NAME 2FGL & NAME XRT & WISE NAME \\
\hline
2FGLJ0031.0+0724 & \textbf{SWXRTJ003119.9+072452} & WISEJ003119.70+072453.6 \\
\hline
2FGLJ0048.8-6347 & SWXRTJ004800.6-634956 & WISEJ004800.63-634951.2\\
\hline
2FGLJ0103.8+1324 & SWXRTJ010414.0+132427 & WISEJ010413.77+132424.5 \\
\hline
2FGLJ0200.4-4105 & SWXRTJ020020.9-410937 & WISEJ020020.94-410935.6 \\
\hline
2FGLJ0221.2+2516 & \textbf{SWXRTJ022051.5+250930} & WISEJ022051.24+250927.6 \\
\hline
2FGLJ0353.2+5653 & \textbf{SWXRTJ035309.5+565429} & WISEJ035309.54+565430.8 \\
\hline
2FGLJ0420.9-3743 & \textbf{SWXRTJ042025.5-374445} & WISEJ042025.09-374445.0 \\
\hline
2FGLJ0427.2-6705 & SWXRTJ042646.3-665954 & WISEJ042646.88-665955.8 \\
\hline
2FGLJ0540.1-7554 & SWXRTJ054112.1-760249 & WISEJ054111.58-760246.1 \\
\hline
2FGLJ0737.1-3235 & SWXRTJ073739.2-323255 & WISEJ073738.91-323256.2 \\
\hline
2FGLJ0737.5-8246 & \textbf{SWXRTJ073706.3-824836} & WISEJ073706.06-824840.2 \\
\hline
2FGLJ0745.5+7910 & SWXRTJ074516.0+791310 & WISEJ074515.65+791312.3 \\
\hline
2FGLJ0746.0-0222 & SWXRTJ074554.9-022430 & WISEJ074554.80-022430.7 \\
\hline
2FGLJ0900.9+6736 & SWXRTJ090123.0+672838 & WISEJ090122.34+672839.9 \\
\hline
2FGLJ1013.6+3434 & SWXRTJ101306.5+343460 & WISEJ101306.10+343501.6 \\
                               & \textbf{SWXRTJ101321.4+343915} & WISEJ101321.17+343912.2 \\
\hline
2FGLJ1033.5-5032 & SWXRTJ103332.0-503531 & WISEJ103332.15-503528.8 \\
\hline
2FGLJ1038.2-2423 & SWXRTJ103748.3-242842 & WISEJ103748.10-242845.5 \\
\hline
2FGLJ1221.4-0633 & SWXRTJ122119.7-063926 & WISEJ122119.72-063927.2 \\
                              & SWXRTJ122154.2-063122 & WISEJ122154.19-063124.1 \\
\hline
2FGLJ1231.3-5112 & SWXRTJ123121.3-511720 & WISEJ123121.67-511717.5 \\
\hline
2FGLJ1328.5-4728 & \textbf{SWXRTJ132840.4-472749} & WISEJ132840.61-472749.2 \\
\hline
2FGLJ1517.2+3645 & SWXRTJ151752.5+364123 & WISEJ151752.12+364125.5 \\
\hline
2FGLJ1704.3+1235 & SWXRTJ170412.0+123658 & WISEJ170412.34+123655.8 \\
\hline
2FGLJ1738.9+8716 & SWXRTJ174142.4+871445 & WISEJ174142.21+871443.6 \\
\hline
2FGLJ1748.8+3418 & SWXRTJ174925.4+341951 & WISEJ174924.98+341951.9 \\
\hline
2FGLJ1842.3+2740 & SWXRTJ184228.3+273345 & WISEJ184228.18+273342.5 \\
\hline
2FGLJ2002.8-2150 & SWXRTJ200224.2-215113 & WISEJ200223.88-215111.6 \\
\hline
2FGLJ2034.7-4201 & SWXRTJ203451.0-420035 & WISEJ203451.08-420038.2 \\
\hline
2FGLJ2115.4+1213 & \textbf{SWXRTJ211521.9+121801} & WISEJ211522.00+121802.8 \\
\hline
2FGLJ2125.0-4632 & SWXRTJ212459.7-464006 & WISEJ212459.90-464008.4 \\
                              & SWXRTJ212515.7-463731 & WISEJ212515.83-463736.8 \\
\hline
2FGLJ2132.5+2605 & SWXRTJ213202.1+260306	& WISEJ213201.89+260306.1 \\
\hline
2FGLJ2228.6-1633 & \textbf{SWXRTJ222830.4-163643} & WISEJ222830.19-163642.8 \\
\hline
2FGLJ2246.3+1549 & \textbf{SWXRTJ224604.9+154437} & WISEJ224604.98+154435.3 \\
\hline
2FGLJ2351.6-7558 & SWXRTJ235115.2-760017 & WISEJ235116.09-760015.5 \\
                               & SWXRTJ235327.5-760018 & WISEJ235328.54-760013.6 \\
\hline
\hline
\end{tabular}
\end{large}
\end{center}
\end{table}

\end{appendix}

\end{document}